\documentclass[11pt,english]{extarticle}
\usepackage{mathpazo}

\usepackage[T1]{fontenc}
\usepackage[utf8]
{inputenc} 
\usepackage[hidelinks]{hyperref}
\usepackage{float}
\usepackage{booktabs}
\usepackage{threeparttable}
\usepackage{algorithm}
\usepackage{algpseudocode} 
\usepackage{multirow}

\usepackage{setspace}
\usepackage{xurl}
\usepackage{microtype}

\usepackage{enumitem}
\setlist[itemize]{topsep=0pt, partopsep=0pt, itemsep=-3pt, left=10pt}
\setlist[enumerate]{topsep=0pt, partopsep=0pt, itemsep=-3pt, left=10pt}
\usepackage[margin=1in]{geometry}
\usepackage{amsmath,amsthm,amssymb, graphicx, multicol, array}
\usepackage[colorinlistoftodos]{todonotes}
\usepackage{natbib}
\usepackage{amsthm}
\theoremstyle{definition}
\newtheorem{assumption}{Assumption}
\newtheorem{definition}{Definition}
\theoremstyle{plain}
\newtheorem{proposition}{Proposition}
\newtheorem{corollary}{Corollary}
\newtheorem{theorem}{Theorem}
\newtheorem{lemma}{Lemma}
\usepackage{graphicx}
\usepackage{subcaption}
\usepackage{algorithm}
\usepackage{algpseudocode}
\usepackage{appendix}
\usepackage{bbm}
\usepackage{authblk}
\usepackage{xurl}

\title{\huge Post-Experiment Decisions: The Dual Adjustments\\ for Rollout and Downstream Optimizations\thanks{The author names are placed in alphabetic order. The authors thank GPT-5.2 for the insightful comments.}}

\author[1]{Guoxing He}
\author[2]{Dan Yang}
\author[3]{Wei Zhang}
\affil[1,2,3]{Faculty of Business and Economics, The University of Hong Kong, Hong Kong}
{
    \makeatletter
    \renewcommand\AB@affilsepx{; \protect\Affilfont}
    \makeatother

    \affil[1]{gxhe@connect.hku.hk}
    \affil[2]{dyanghku@hku.hk}
    \affil[3]{wzhang15@hku.hk}
}
\date{}

\begin{document}
\maketitle
\begin{abstract}

\noindent Firms increasingly use randomized experiments to decide whether to scale up an intervention and, if so, how to re-optimize related operational choices such as inventory, capacity, or pricing. In many settings, experiments are performed on small samples, so the estimated effect of the intervention is uncertain. A common practice is to plug a "significant" estimate of the effect into both (i) the rollout rule and (ii) the downstream optimization. However, this can lead to avoidable losses because the costs of over- versus under-estimating the effect are often asymmetric. The technically ideal approach is to obtain a data-dependent decision rule that minimizes the Bayes risk, but this lacks transparency and requires more computations. We propose Predict–Adjust–Then–Rollout–Optimize (PATRO), a plug-in approach that keeps the standard estimate, but makes data-independent adjustments, respectively, for the two types of decision. We show that the two adjustments can be substitutes or complements and provide an alternating-iteration method to compute the pair. Surprisingly, PATRO performs both in theory and numerically close or equivalent to the Bayes-optimal benchmark, making it a simple, effective way to convert noisy experimental results into better rollout and operational decisions without changing a firm's estimation pipeline or decision models.

\vspace{5mm}
\end{abstract}
\vspace{2cm}
\pagebreak{}

\section{Introduction}

Firms increasingly rely on randomized experiments to guide high-stakes operational choices. While large technology companies run thousands of A/B tests annually, multiunit enterprises (MUEs), such as retail chains, restaurants, and transportation networks, are also adopting experimentation to evaluate local changes (e.g., store layout, pricing, service innovations) before scaling them across a network. In these settings, experimentation is costly and often limited to a small number of units; therefore, post-experiment decisions must be made under substantial estimation uncertainty. Importantly, the decision problem is rarely limited to a binary ``ship/no-ship'' choice. After an experiment, firms typically face a \emph{two-stage} decision: whether to roll out the intervention, and conditional on rollout, how to re-optimize downstream operational choices at each unit (e.g., inventory, capacity, staffing, or pricing). This paper studies how to translate noisy causal evidence into high-quality decisions in such two-stage environments.

Here are two examples that illustrate the challenge.
\begin{enumerate}
    \item A retailer pilots self-checkout kiosks in a small set of stores. After estimating the demand impact, the firm must decide whether to deploy kiosks chain-wide and, if deployed, how each store should adjust inventory and pricing to reflect the new demand level.
    \item A restaurant chain pilots tablet ordering to reduce dining time and increase table turnover. After estimating the reduction in dining time, the firm must decide whether to roll out the technology and, if so, how to adjust table capacity across outlets.
\end{enumerate}In both cases, the rollout decision interacts with downstream optimization: overestimating the effect may trigger an unprofitable rollout and overly aggressive operational investments, while underestimating may forgo a profitable rollout and induce overly conservative operations. Because these losses are typically asymmetric, a decision rule that is optimal for statistical accuracy (i.e., minimizing certain empirical losses) need not be optimal for economic performance.

A common approach follows the \emph{Predict-Then-Optimize} (PTO) paradigm: estimate the treatment effect and plug the point estimate (often the posterior mean or the sample mean difference) into downstream optimization models. However, PTO can be systematically suboptimal in our setting for two reasons. First, optimization amplifies estimation noise (the optimizer's curse), and over- versus under-estimation can create asymmetric regret in the downstream objective. Second, the presence of a binary rollout decision coupled with continuous operational choices makes the post-experiment decision problem nonlinear and nonconvex; as a result, standard decision-aware estimation methods and one-stage adjustment logic do not directly apply.
\begin{figure}
    \centering
    \includegraphics[scale=0.6]{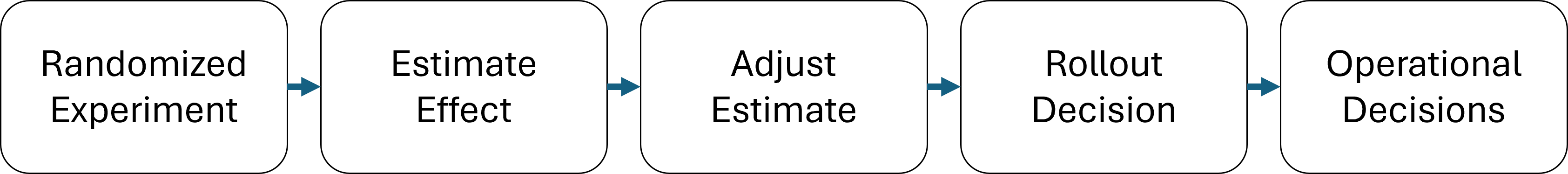}
    \caption{The PATRO Workflow}
\end{figure}

We propose a simple and tractable alternative: \emph{Predict-Adjust-Then-Rollout-Optimize} (PATRO). We retain standard causal estimation in the prediction step but deliberately \emph{adjust} the effect estimate before it enters the rollout rule and/or the downstream optimization. The adjustments are chosen to minimize \emph{prior expected regret} (Bayes risk) induced by post-experiment decisions under finite-sample uncertainty. We focus on a quantile-based adjustment rule within a Bayesian framework. Under a Gaussian posterior, the PTO estimator equals the posterior mean (and median), corresponding to the $0.5$-quantile. PATRO instead selects an optimal posterior quantile, equivalently an additive shift from the posterior mean, to reflect the asymmetry of decision losses as well as the expected size of estimation error. Crucially, because rollout and operational decisions occur in stages, we allow \emph{two distinct adjustments}: one for each stage.

Our analysis provides three main insights. First, the optimal adjustments can be either positive or negative relative to the posterior mean, depending on the economic structure of the downstream objective. Intuitively, when the payoff as a function of the true effect is concave, downside losses dominate upside gains, and the rollout rule should be more conservative; when it is convex, the reverse is true. Second, the two adjustments are generally not independent. In a two-stage environment, the rollout adjustment and the operational adjustment can be \emph{substitutes} (adjusting one reduces the magnitude of the other) or \emph{complements} (adjusting one increases the magnitude of the other), meaning that the rollout and subsequent operational decisions cannot be dealt with independently. Third, and surprisingly, while PATRO's decisions are restricted to plugging fixed posterior quantiles into oracle decision rules, the performance gap in terms of prior expected regret between PATRO and the Bayes optimal rule (minimizing the posterior expected regret with a general decision rule) is negligible or even zero in many cases.

Methodologically, we formulate a unified Bayesian causal decision-making model with (i) experimental estimation of an unobservable treatment effect, (ii) a binary implementation decision, and (iii) a concave downstream optimization problem whose value depends on the true effect and the effect estimate used for decision-making. We characterize regret into three components---Type I error (implementing when the effect is negative), Type II error (not implementing when the effect is positive), and operational regret (mis-calibration of downstream decisions conditional on correct implementation). Within this framework, we derive first-order optimality conditions for each adjustment, show that optimal adjustments vanish at rate $O(n^{-1})$ as the sample and firm size $n$ increases, and establish conditions under which dual adjustments decouple or interact. We also provide a simple alternating-iteration procedure to compute the dual adjustments and prove its convergence under mild regularity. We prove the equivalence between PATRO and the Bayes optimal rule under a non-trivial condition.

This paper contributes to two research streams. Relative to the PTO and post-estimation adjustment literature, we extend adjustment logic from single continuous decisions to a two-stage binary--continuous structure and show that optimal conservatism/aggressiveness depends on curvature and cross-curvature properties of the downstream value function. Relative to causal decision-making and policy learning, we go beyond binary treatment assignment by integrating rollout with endogenous operational scaling decisions that depend on the estimated effect.

The remainder of the paper is organized as follows. Section~\ref{sec:model} presents the Bayesian estimation model and the two-stage decision framework. Section~\ref{sec:single_adjustment} analyzes single adjustments for rollout and for downstream operations. Section~\ref{sec:double} studies dual adjustments, characterizes when the adjustments are substitutes versus complements, and presents a convergent algorithm for computing them. Section~\ref{sec:benchmark} compares PATRO against the Bayes optimal rule theoretically. Section~\ref{sec:numerical} provides numerical examples from inventory, service operations, and pricing to illustrate the diversity of optimal adjustment patterns across environments. Section \ref{sec:conclusion} concludes the paper. All mathematical details and proofs are provided in the Appendix.

\section{Literature Review}

Our work bridges two converging research streams: (i) the predict-then-optimize paradigm and post-estimation adjustment in operations, and (ii) causal decision-making, policy learning, and experimentation. We contribute a unified framework that jointly optimizes a binary implementation (rollout) decision and continuous downstream operational decisions under estimation uncertainty from small-sample experiments. More broadly, our setting differs from much of the prior literature along three dimensions: \emph{managerial context} (single unit versus multiple units), \emph{data structure} (observable versus unobservable ground truth), and \emph{decision structure} (optimization with versus without a rollout decision). Our analysis explicitly incorporates multiunit decision-making, an unobservable causal parameter (the treatment effect), and the interaction between a rollout threshold and endogenous operational decisions.

\subsection{Predict-then-Optimize Framework and Post-Estimation Adjustment}

The predict-then-optimize (PTO) paradigm separates statistical estimation from downstream decision-making. This modular workflow is widely used in practice and is theoretically sound in certain contexts \citep{hu2022fast}. However, in many other contexts, it can be vulnerable to the \emph{optimizer's curse} \citep{smith2006optimizer}: estimation noise is amplified by optimization, leading to systematically biased decisions and inferior realized performance.

One response integrates prediction and optimization at the estimation stage. The Smart Predict-then-Optimize (SPO) framework of \citet{elmachtoub2022smart} replaces standard prediction losses with decision-aware objectives that directly minimize downstream regret. SPO exemplifies the emerging paradigm of decision-focused learning (DFL), which embeds constrained optimization problems directly into the machine learning training pipeline, enabling models to be jointly optimized with respect to the quality of resulting decisions rather than intermediate prediction metrics (see \cite{ mandi2024decision} for a comprehensive review).

A complementary literature, exemplified by the Operational Data Analytics (ODA) framework \citep{feng2022developing}, characterizes structural environments in which specific operational statistics yield optimal downstream decisions, obviating the need for decision‑aware estimation. The Operational Data Analytics (ODA) framework characterizes structural environments---such as homogeneity or separability---under which particular operational statistics induce uniformly optimal decision rules. Applications include consumer choice estimation \citep{feng2022consumer}, service system design \citep{feng2025operational}, and pricing and inventory models \citep{chu2025solving}.

Distinct from both decision‑aware estimation and structural optimality results, a new stream of research studies post‑estimation adjustment strategies that preserve standard estimators but modify how estimates enter downstream decision problems. This approach recognizes that estimation noise interacts asymmetrically with optimization and that deliberately biasing estimates can reduce expected decision loss.
Our work is most closely related to \citet{albert2025post}, who develop a curvature‑based post‑estimation adjustment for data‑driven pricing. They show that asymmetry in the profit function around the optimum induces a bias–variance trade‑off in the decision space, implying that biased estimators can outperform unbiased ones in terms of realized profit. We build directly on this insight by extending curvature‑based adjustment to a causal decision‑making setting with both binary and continuous decisions. In addition, our Bayesian optimization framework offers new insights about the impact of "2-D" skewness of the surrogate reward function.

We depart from this literature by analyzing a two-stage decision structure commonly encountered in practice: a binary implementation decision (whether to act) followed by a continuous operational decision (how much to invest). We show that the interaction between these two stages fundamentally alters the nature of optimal adjustment. In contrast to single-stage settings—where adjusting decisions after estimation serves a similar purpose as shrinking estimates toward conservative values—we characterize conditions under which binary implementation adjustments and continuous operational adjustments could act as complements rather than substitutes.

\subsection{Causal Decision-Making, Policy Learning, and Experimentation}
Recent advances in causal inference have shifted attention from estimation accuracy to decision quality under uncertainty. A central insight is that estimators optimized for effect estimation need not be optimal for causal decision-making: when the objective is to choose whom to treat, what matters is accurate treatment assignment rather than accurate effect-size estimation \citep{fernandez2022causal}. Indeed, biased outcome prediction models can outperform unbiased treatment effect estimators for causal classification, particularly when data for estimating counterfactuals are limited \citep{fernandez2022causal_classification}. Similar to this insight, the policy learning literature directly optimizes treatment assignment rules for welfare or regret, rather than first estimating treatment effects and then deriving policies \citep{athey2021policy, kitagawa2018should}.

While these advances clarify how estimation should adapt to decision objectives, the focus remains largely on binary implementation decisions—whether to treat or not. A parallel literature on experimentation and rollout shares this emphasis. The Test \& Roll framework \citep{feit2019test} reframes A/B testing as a profit-maximization problem, deriving optimal test sizes by trading off the opportunity cost of experimentation against the risk of deploying a suboptimal treatment. Related work optimizes experimental design for precision \citep{bhat2020near} or accommodates fat-tailed treatment effect distributions \citep{azevedo2020ab}. More recently, \cite{peng2025synthesizing} develop a data-pooling treatment roll-out (DPTR) framework that reduces estimation variability by pooling data across multiple experiments rather than analyzing each in isolation.
However, these frameworks typically treat post-test deployment as automatic: once a winning variant is selected, implementation proceeds at a fixed or exogenous scale.

In many operations management settings, decisions involve not only whether to implement a treatment but also how much to invest conditional on implementation—for example, in inventory, capacity, or budget. Small-sample experiments leave substantial estimation uncertainty that propagates into both decisions. We address this by jointly optimizing the binary implementation threshold and the continuous operational decisions, characterizing the conditions under which regret minimization requires adjusting one alone or both.

\section{The Model}\label{sec:model}
We follow the Predict-Adjust-Then-Rollout-Optimize (PATRO) workflow described earlier. First, we describe the estimation procedure of the treatment effect (the "P" step) as well as the statistical properties of the estimate in the Bayesian framework. Second, we introduce the decision making framework in the "RO" step. We will explain the "A" step in the next section.

\subsection{Effect Estimation in the Bayesian Framework}\label{sec:bayesian_model}

We adopt the potential outcome framework \citep{rubin1974estimating,rubin1978bayesian}. Consider a population of $n$ units indexed by $i = 1,\dots,n$. Each unit can be exposed to a binary treatment $W_i \in \{0,1\}$. For each unit $i$, we define two potential outcomes:
$Y_i^1$ and $Y_i^0$, corresponding to treatment ($W_i=1$)
and control ($W_i=0$), respectively. We consider an additive treatment effect and a simple linear structure for the potential outcomes:
\begin{equation}
\label{eq:potential}
Y_i^0 = b + \varepsilon_i,\quad\text{and}\quad Y_i^1 = b + \tau + \varepsilon_i,
\end{equation}
where $b$ is a known baseline mean level (common to all units), $\tau$ is the constant treatment effect of interest across units, $\varepsilon_i$ is an idiosyncratic error term. This model implies that $Y_i^1 - Y_i^0 = \tau$ for all $i = 1,\dots,n$.
We assume that the error terms are independent and identically distributed (i.i.d.): $\varepsilon_i \sim \mathcal{N}(0, \sigma_\varepsilon^2)$ with known variance $\sigma_\varepsilon^2$. We further assume that $\{\varepsilon_i\}_{i=1}^n$ are independent of $\tau$ and of the treatment assignment (defined below). Note that we can estimate $b$ and $\sigma_\varepsilon$ with historical data to sufficient accuracy, so we assume they are known.

We maintain the standard Stable Unit Treatment Value Assumption
(SUTVA) \citep{rubin1978bayesian}, assuming no interference between
units and well-defined potential outcomes.
Under SUTVA, the observed outcome satisfies the consistency condition
$
Y_i = W_i Y_i^1 + (1-W_i) Y_i^0.
$
Substituting the potential outcome model in \eqref{eq:potential}, for a realization $(y_i,w_i)$ of $(Y_i,W_i)$, we have
\begin{equation}
\label{eq:obs_model}
y_i = b + \tau w_i + \varepsilon_i, \quad i=1,\dots,n.
\end{equation}

To ensure the validity of causal inference, we stipulate that the data are generated via a completely randomized experiment. This design guarantees that the treatment assignment is unconfounded, as formally defined below:
\begin{assumption}
\label{ass:CRE}
Let
$
\mathbf Y
=
\big((Y_1^1,Y_1^0),\dots,(Y_n^1,Y_n^0)\big)
$
denote the vector of potential outcomes; and let
$
\mathbf W=(W_1,\dots,W_n)\in\{0,1\}^n
$ denote the treatment assignment vector, and satisfy the following conditions:
    \begin{enumerate}
        \item \textbf{Exogeneity:} The assignment vector $\mathbf{W}$ is statistically independent of the treatment effect $\tau$ and of the error terms $\{\varepsilon_i\}_{i=1}^n$; hence, $\mathbf W$ is independent of the potential outcomes
    $\mathbf Y$.
        \item \textbf{Fixed Sample Sizes:} The vector $\mathbf{W}$ is drawn uniformly at random from the set of all possible assignment vectors with exactly $n_1=\gamma (1+\gamma)^{-1}n$ treated units, where $\gamma\in(0,1)$ is the fixed ratio between the treatment and control group sizes.
    \end{enumerate}
\end{assumption}
Assumption~\ref{ass:CRE} implies that the assignment mechanism is \emph{ignorable} for posterior inference about $\tau$ in the Bayesian framework. Therefore, in what follows, all the calculation of the likelihood, posterior, and other relevant quantities are conditioning upon the realization of assignment $\mathbf{W}$.

The fixed sample size design ensures that the numbers of treated and control units, $n_1$ and $n_0$, are deterministic rather than random. An alternative is Bernoulli randomization, where units are independently assigned to treatment with a fixed probability. While the two schemes are asymptotically equivalent for many purposes, fixed treatment and control group sizes are often preferred in practice (e.g., due to budget or capacity constraints) and simplify the exposition here.

Finally, we adopt a Bayesian perspective on the treatment effect $\tau$. Since the data generating process is normal, we choose the conjugate prior of the normal distribution for the treatment effect, $\tau \sim \mathcal{N}(m_0, v_0)$, where the mean $m_0$ and variance $v_0 > 0$ are selected to reflect prior knowledge. Again, $\tau$ is independent of the noise $\{\varepsilon_i\}$ and assignment $\mathbf{W}$.
\vspace{-0.5em}
\paragraph{Group means and sampling distribution.}
For a realization $\mathbf s=\{(y_i,w_i)\}_{i=1}^n$ of the experimental data sample
$\mathbf S$, let $I_1 = \{i \mid w_i = 1\}$ and $I_0 = \{i \mid w_i = 0\}$ denote the sets of indices for the treatment and control groups, respectively. We define the sample means for each group as:
\begin{equation*}
    \bar{y}^1 = \frac{1}{n_1}\sum_{i \in I_1} y_i \quad\text{and}\quad
    \bar{y}^0 = \frac{1}{n_0}\sum_{i \in I_0} y_i.
\end{equation*}
We refer to the difference in sample means as the \textbf{Naive Estimator} of the treatment effect:
\[
    \hat{\tau}_{ne} \equiv \bar{y}^1 - \bar{y}^0
    = (b + \tau + \bar{\varepsilon}^1) - (b + \bar{\varepsilon}^0)
    = \tau + (\bar{\varepsilon}^1 - \bar{\varepsilon}^0),
\]
where $\bar{\varepsilon}^1$ and $\bar{\varepsilon}^0$ represent the averages of the independent error terms for each group. Since errors are normally distributed, the naive estimator also follows a normal distribution conditional on $\tau$:
\[
    \hat{\tau}_{ne} \mid \tau
    ~\sim~
    N\left( \tau, \,\, \sigma_\varepsilon^2 \left[\frac{1}{n_1} + \frac{1}{n_0}\right] \right).
\]

\paragraph{Likelihood and Prior.} For a realization $\mathbf s=\{(y_i,w_i)\}_{i=1}^n$ of the sample
$\mathbf S$. It is easy to see that $\bar{y}^1 - \bar{y}^0$ is a sufficient statistic for $\tau$. Based on the sampling distribution above, the likelihood function of $\tau$ with the realized data $\mathbf s$ is
\[
p(\mathbf s\mid\tau)
~\propto~
\exp\!\left\{
-\frac{
\big[(\bar y^1-\bar y^0)-\tau\big]^2
}{
2\sigma_\varepsilon^2\left(\frac{1}{n_1}+\frac{1}{n_0}\right)
}
\right\}.
\]
The prior distribution of $\tau$ is specified as $\tau \sim \mathcal{N}(m_0, v_0)$, with density:
\[
    f(\tau)
    ~\propto~
    \exp\left\{
    -\frac{(\tau - m_0)^2}{2v_0}
    \right\}.
\]
Applying the Bayes' theorem, the posterior density of $\tau$ is:
\[
p(\tau\mid \mathbf S=\mathbf s)
~\propto~
p(\mathbf s\mid\tau)\,f(\tau)
~\propto~
\exp\!\left\{
-\frac{(\tau-m_0)^2}{2v_0}
-
\frac{
\big[(\bar y^1-\bar y^0)-\tau\big]^2
}{
2\sigma_\varepsilon^2\left(\frac{1}{n_1}+\frac{1}{n_0}\right)
}
\right\}.
\]

\paragraph{Posterior.} Since the exponent in the likelihood-prior product is quadratic in~$\tau$, the posterior distribution is also Normal:
\[
\tau \mid \mathbf S=\mathbf s
~\sim~
\mathcal N\!\big(\tilde m(\mathbf s),\,\tilde v\big),
\]
where
\begin{equation}\label{eq:vtilde_mtilde}
\tilde{v}^{-1}
~=~
v_0^{-1}
~+~
\frac{n\cdot \gamma}{\sigma_\varepsilon^2(1+\gamma)^2} \quad\text{and}\quad
\tilde{m}
~=~
\frac{\tilde{v}}{v_0}\cdot m_0~+~\frac{n\cdot \gamma\cdot \tilde{v}}{\sigma_\varepsilon^2(1+\gamma)^2}\cdot\hat{\tau}_{ne}.
\end{equation}
Here, we have used the sample size ratio $\gamma=n_1/n_0$ and substituted \(1/n_1 + 1/n_0 = (1+\gamma)^2/(\gamma n)\). Throughout, we write $\tau\mid \mathbf s$ as shorthand for
$\tau\mid \mathbf S=\mathbf s$.

Under the PTO paradigm, the goal of the ``P'' step is to minimize the posterior expected squared error, leading to the \textbf{Predict-then-Optimize (PTO)} estimator
\begin{equation}
\hat{\tau}_{pto}(\mathbf s)
=
\operatorname*{arg\,min}_{a\in\mathbb R}
\;
\mathbb E\!\left[
(a-\tau)^2
\,\middle|\,
\mathbf s
\right]
=
\tilde m(\mathbf s).
\end{equation}

Note that the naive estimator $\hat{\tau}_{ne}$ equals the PTO estimator $\hat{\tau}_{pto}$ when the prior belief of $\tau$ satisfies $v_0=\infty$ or when the sample size $n=\infty$. In other words, when there is no prior knowledge about $\tau$ or there is sufficient empirical evidence, we completely rely on the naive estimator.

Next, we establish the following theoretical results characterizing the relationship between the true treatment effect $\tau$ and the PTO estimator $\tilde{m}$.

\begin{lemma}\label{lemma_tildem_tau}
    The PTO estimator $\tilde{m}$ satisfies the following properties:
    \begin{enumerate}
        \item The pair $(\tau, \tilde{m})^\top$ follows a bivariate normal distribution:
        \[
        \begin{pmatrix} \tau \\ \tilde{m} \end{pmatrix} \sim \mathcal{N} \left(
        \begin{pmatrix} m_0 \\ m_0 \end{pmatrix},
        \begin{pmatrix} v_0 & v_m \\ v_m & v_m \end{pmatrix}
        \right),
        \]
        with density function $f(\tau, \tilde{m})=$
\begin{equation}\label{eq:ftautildem}
         \frac{1}{2\pi \sqrt{v_m(v_0 - v_m)}} \exp\left( -\frac{v_m(\tau - m_0)^2 - 2v_m(\tau - m_0)(\tilde{m} - m_0) + v(\tilde{m} - m_0)^2}{2v_m(v_0 - v_m)} \right),
        \end{equation}
        where $v_m = v_0^2 / (v_0 + \sigma_\varepsilon^2 (1+\gamma)^2 / (\gamma n))$, and $\tilde{v}$ and $\tilde{m}$ are as defined in \eqref{eq:vtilde_mtilde}.

        \item The estimation error defined as $\epsilon_e \equiv \tilde{m} - \tau$ is independent of $\tilde{m}$ and follows $\epsilon_e \sim \mathcal{N}(0, \tilde{v})$.
    \end{enumerate}
\end{lemma}

\subsection{The Decision-Making Framework}\label{sub:decision-framework}

The firm makes two types of decision: A binary rollout decision $D \in \{0,1\}$ that pertains to whether to implement the new treatment ($D=1$) across all units or to stick to the status quo ($D=0$); Operational decisions at the downstream unit-level $\textbf{u}=(u_1,...,u_n)$, where $u_i \in \mathcal{U}$ applies for each unit $i = 1,\dots,n$. In practice, the rollout scale may differ from the sample size in the experiment, but they should grow proportionally; hence, for simplicity, we assume that they are equal.

Let $G_i(u_i\mid\tau)$ denote the objective function of the optimization problem (e.g., payoff, profit) faced by unit $i$ when the ground-truth treatment effect is $\tau$ and the decision is $u_i$. Let $c$ be the fixed cost of rolling out the treatment, and $\bar{u}_i$ the status quo optimal decision for unit $i$ when $\tau = 0$. Define the baseline payoff under the status quo as $G_0 \equiv \sum_{i=1}^n G_i(\bar{u}_i|0)$, which is a constant that is independent of the decisions.
Given the true $\tau$, a generic decision profile $(D, \textbf{u})$ yields total payoff
\begin{equation}
   \label{eq:Gamma_function} \Gamma(D, \textbf{u} \mid \tau) = D \cdot \sum_{i=1}^n G_i(u_i\mid\tau) + (1 - D) \cdot G_0 - c \cdot D.
\end{equation}
Without loss of generality, we normalize $c$ to zero in our subsequent analysis.

If the true $\tau$ were known, the firm would choose unit-level decisions by solving, for each $i$,
\begin{equation}\label{eq:u}
    u_i^{\#}(\tau) \equiv \arg\max_{u \in \mathcal{U}} G_i(u\mid\tau),
\end{equation}
and then select the rollout decision
\begin{equation}
D^{\#}(\tau)
\equiv
\mathbb I\!\left\{
\sum_{i=1}^{n} G_i\!\big(u_i^{\#}(\tau)\mid\tau\big)
> G_0
\right\}.
\end{equation}
We impose (and later formalize) the assumption that the oracle payoff conditional on rollout, $\sum_{i=1}^n G_i\bigl(u_i^{\#}(\tau)\mid\tau\bigr)$, is weakly increasing in $\tau$ for our maximization problem. Under this monotonicity and the normalization of $c=0$, the oracle rollout decision takes the natural threshold form
\begin{equation}
\label{eq:D}
  D^{\#}(\tau)=\mathbb{I}\{\tau>0\},
\end{equation}
where $\mathbb{I}\{\cdot\}$ is the binary indicator function. That is, if the true treatment effect is beneficial ($\tau>0$), the firm should roll out; otherwise it should not.

In practice, $\tau$ is unknown and the firm uses an estimate $\hat{\tau}$ (from the “P” step). A natural policy is: For rollout, plug $\hat{\tau}$ into the oracle rule and use $D^{\#}(\hat{\tau})$; For unit-level decisions, plug $\hat{\tau}$ into the oracle unit rules and use $u_i^{\#}(\hat{\tau})$.
The resulting net payoff from rollout (relative to the status quo), referred to as the \textbf{Surrogate Net Reward (SNR) Function}, is written as
\begin{equation}
\label{eq:SNR_function}    \Pi(\hat{\tau} \mid \tau) \equiv \sum_{i=1}^n G_i\bigl(u_i^{\#}(\hat{\tau}) \mid \tau\bigr)-G_0,
\end{equation}
and the resulting payoff is characterized by the \textbf{Surrogate Reward Function}:
\begin{equation}
    \tilde{\Pi}(\hat{\tau} \mid \tau) \equiv D^{\#}(\hat{\tau}) \cdot \Pi(\hat{\tau} \mid \tau) +  G_0.
\end{equation}

Note that this surrogate reward depends on the ground-truth state $\tau$ and the estimate $\hat{\tau}$ provided by the ``P'' step. In its current form, it assumes that both decisions, $D^{\#}(\hat{\tau})$ and $u_i^{\#}(\hat{\tau})$, are based on the same estimate. However, the rollout decision and the downstream unit decisions are made in two stages, which allows the firm to make adjustments for the two decisions separately. Hence, we use $\hat{\tau}^r$ and $\hat{\tau}^o$, respectively, to represent the \textbf{\textit{adjusted estimates}} for the rollout decision (r) and the downstream operational decisions (o), and we rewrite the surrogate reward as
\begin{equation}
    \tilde{\Pi}(\hat{\tau}^r,\hat{\tau}^o \mid \tau)\equiv D^{\#}(\hat{\tau}^r)\cdot\Pi(\hat{\tau}^o\mid\tau)+G_0.
\end{equation}

We now introduce the technical assumptions for the SNR function $\Pi(\cdot|\cdot)$. For notation simplicity, let $\Pi^{(\alpha_1,\alpha_2)}(\tau_1\mid\tau_2)$ denote the partial derivatives of order $\alpha_1$ and $\alpha_2$ with respect to the first and second arguments, respectively, evaluated at the point $(\tau_1, \tau_2)$:
$$\Pi^{(\alpha_1,\alpha_2)}(\tau_1\mid\tau_2) \equiv \frac{\partial^{\alpha_1+\alpha_2} \Pi(\hat{\tau}^o\mid\tau)}{(\partial \hat{\tau}^o)^{\alpha_1} (\partial \tau)^{\alpha_2}} \bigg|_{\hat{\tau}^o=\tau_1, \tau=\tau_2}.$$
\begin{assumption}\label{ass_pitauhat}
    For the surrogate net reward (SNR) $\Pi(\hat{\tau}^o\mid\tau)$, we assume
    \begin{enumerate}
        \item[(i)] \textbf{Smoothness and Optimality at Truth:} The SNR function $\Pi(\hat{\tau}^o\mid\tau)$ is at least three times continuously differentiable with respect to $\tau$ and $\hat{\tau}^o$ ($C^3$). Given perfect estimate $\hat{\tau}^o=\tau$, it satisfies: $$\Pi^{(1,0)}(\tau\mid\tau) = 0, \quad \Pi^{(2,0)}(\tau\mid\tau) < 0.$$
        \item[(ii)] \textbf{Expected Concavity in $\hat{\tau}^o$:} For any $r \in \mathbb{R}$ and any $\delta^o \in \mathbb{R}$, if we set $\hat{\tau}^o = \tilde{m} + \delta^o$,
        \[
        \mathbb{E}_{\tau, \tilde{m} > -r}\left[\Pi^{(2,0)}(\hat{\tau}^o \mid \tau)\right] < 0.
        \]
         \item[(iii)] \textbf{Monotonicity of the oracle SNR:} $\Pi(\tau\mid\tau)$ weakly increases in $\tau$.

\item[(iv)]
    \textbf{Polynomial Growth:} There exist constants $C > 0$ and $K \ge 0$ such that for all $\hat{\tau}^o$ and $\tau$, and for all $i,j \in \{0,1,2,3\}$ with $i+j \le 3$:
    \[
    \left| \Pi^{(i,j)}(\hat{\tau}^o\mid\tau) \right| \le n \cdot C(1 + |\hat{\tau}^o|^K + \mid\tau|^K).
    \]
  \item[(v)] \textbf{Posterior Expected SNR Function Regularity:} Define posterior expected surrogate net reward function $V : \mathbb{R} \to \mathbb{R}$ by
\begin{equation}
\label{eq:value_function}
    V(m,\delta) \equiv \mathbb{E}_{\tau \mid \tilde{m} = m} \left[ \Pi(\hat{\tau}^o \mid \tau) \right],
\end{equation}
where $\hat{\tau}^o = \tilde m + \delta$ for a given adjustment $\delta$. The posterior expected SNR function satisfies the following conditions:
\begin{enumerate}
    \item \textbf{Strict Monotonicity:}
    $V'(m,\delta) = \mathbb{E}_{\tau \mid \tilde{m} = m} \left[ \Pi^{(1,0)}(\hat{\tau}^o \mid \tau) + \Pi^{(0,1)}(\hat{\tau}^o \mid \tau) \right]>0
    $ for all $m$.
    \item \textbf{Boundary Conditions:}
    $
    \lim_{m \to -\infty} V(m,\delta) < 0 \  \text{and} \ \lim_{m \to +\infty} V(m,\delta) > 0.
    $
\end{enumerate}
\end{enumerate}
\end{assumption}
\noindent\textit{Remarks.} Assumption \ref{ass_pitauhat}(i) requires that the downstream operational decision problems are concave and are unconstrained when optimal decisions are chosen. Assumption \ref{ass_pitauhat} (ii) guarantees the existence and uniqueness of the optimal operational decision adjustment, and it requires that the SNR is concave in the estimate on average with respect to a set of conditional beliefs. Assumption \ref{ass_pitauhat} (iii) is intuitive, stating that the "oracle net payoff" (net payoff under perfect information) increases with the true treatment effect. Assumption \ref{ass_pitauhat} (iv) satisfy the sufficient conditions for the Leibniz Integral Rule, enabling the exchange of the order of differentiation and expectation. Finally, Assumption \ref{ass_pitauhat} (v) guarantees the existence and uniqueness of the optimal rollout decision adjustment. We admit that these assumptions make the propose method less general, but as we will show later these assumptions can be satisfied by many different problems.

\vspace{-1em}
\paragraph{Regret.}
To assist our analysis, it is useful to define the \textbf{Regret} as the loss relative to the maximum payoff achieved with perfect knowledge of $\tau$. Formally, we define the regret as
\begin{equation}
    R(\hat{\tau}^r,\hat{\tau}^o \mid \tau) = \tilde \Pi(\tau,\tau \mid \tau) - \tilde\Pi(\hat{\tau}^r,\hat{\tau}^o \mid \tau).
\end{equation}
Decomposing this by scenario yields:
\begin{equation}
\label{eq:regret-Pi}
    R(\hat{\tau}^r,\hat{\tau}^o\mid \tau) = \begin{cases}
        \Pi(\tau \mid \tau) & \text{if } D^{\#}(\tau)=1, D^{\#}(\hat{\tau}^r)=0 \quad (\text{Type II Error}), \\
        -\Pi(\hat{\tau}^o \mid \tau) & \text{if } D^{\#}(\tau)=0, D^{\#}(\hat{\tau}^r)=1 \quad (\text{Type I Error}), \\
        \Pi(\tau \mid \tau) - \Pi(\hat{\tau}^o \mid \tau) & \text{if } D^{\#}(\tau)=1, D^{\#}(\hat{\tau}^r)=1 \quad (\text{operational regret}), \\
        0 & \text{otherwise.}
    \end{cases}
\end{equation}
The first two terms represent classical "wrong-decision" regrets caused by Type II Error and Type I Error, respectively. The third term captures the \textit{operational regret} that arises when the treatment is correctly implemented, but unit operational decisions are suboptimal because they are tuned to $\hat{\tau}^o$ rather than $\tau$. Note that the difference in regret between $D^{\#}(\hat{\tau}^r)=0$ and $D^{\#}(\hat{\tau}^r)=1$ is the driver of adjustment for the rollout decision. In addition, only the Type I Error and operational regret are related to the operational decisions.

\section{Decision-Making Adjustments}\label{sec:single_adjustment}

In this section, we analyze two single adjustment scenarios: adjusting the effect estimate only for the rollout decision or for the downstream unit decision.

\subsection{Adjustment for the Rollout Decision}\label{sec:Rollout_Adjustment}

Given the rollout decision rule (i.e., \eqref{eq:D}) in the previous section, we now refine the decision to account for two key operational realities. First, decision makers often face asymmetric regret: the economic consequences of erroneously implementing a treatment (false positive) may differ substantially from those of failing to implement a beneficial treatment (false negative). Second, estimation error creates systematic bias in the decision, leading to regret. These considerations motivate us to systematically bias the effect estimate before passing it to the rollout decision maker, who then plugs it into the rollout decision model. Here, we assume that the rollout decision maker knows that the effect estimate is not adjusted for the downstream unit decisions (i.e., $\hat{\tau}^o=\tilde{m}$).

\paragraph{Quantile-Based Adjustment Rule}

Because the posterior distribution is normal, the PTO estimator (i.e., the posterior mean) is also the median (quantile $q=0.5$). If we make the rollout decision based on the comparison between the posterior mean against 0 (with $c=0$), we treat false positives and false negatives in a symmetric way. To address the above-mentioned challenges, we consider a rollout decision rule based on an optimal posterior quantile of the treatment effect. Specifically, we set $\hat{\tau}^r$ to the $q$-th quantile on the posterior distribution. With $q\neq 0.5$, we aim to create a safety margin aligned with the specific risk profile of the problem. Formally, let
$
F^{-1}_{\tau \mid \mathbf s}
$
denote the inverse cumulative distribution function (quantile function)
of $\tau$ conditional on the observed data $\mathbf S=\mathbf s$.
The oracle rollout rule can then be written as
\begin{equation}
\label{eq:oracle_quantile_rule}
D^{\#}\!\left(\hat\tau^{r}(\mathbf s)\right)
=
\mathbb I\!\left\{
F^{-1}_{\tau \mid \mathbf s}(q) > 0
\right\}.
\end{equation}

This rule dictates that the decision maker implements the treatment if and only if the $q$-th quantile of the treatment effect is positive. Under the Gaussian posterior $\tau \mid \mathbf{s} \sim \mathcal{N}(\tilde{m}, \tilde{v})$, the quantile function is given explicitly by $F^{-1}_{\tau \mid \mathbf{s}}(q) = \tilde{m} + \sqrt{\tilde{v}}z_q$, where $z_q=\Phi^{-1}(q)$ is the $q$-th quantile of the standard normal distribution. Consequently, the adjusted estimator for rollout decision is given by
\begin{equation}
    \label{eq:hat_tau_r_n}
    \hat{\tau}^r=\tilde{m} + \delta_n^r,
\end{equation}
wherein $\delta_n^r \equiv \sqrt{\tilde{v}}z_q$ is the adjustment from $\tilde{m}$. Later, we will show that $\delta_n^r$ converges to 0 at the rate of $O(\tilde{v})$. Note that $\tilde{v}$ depends on the total sample size $n$ deterministically and converges to 0 at the rate of $O(n^{-1})$ as $n\rightarrow +\infty$. Hence, the optimal $q$ is also approaching 0.5 at the rate of $O(n^{-1/2})$. This (convergence) confirms the asymptotic optimality of the PTO estimator, but the convergence rate reveals a significant gap between PTO and our new approach under a small sample.

\paragraph{The Optimal Adjustment}
We focus on the optimal $q$ (or equivalently $\delta_n^r$) that minimizes the \textit{prior expected regret} (before the experiment data and $\tilde{m}$ are realized), defined as
\begin{equation}
    \mathcal{R}^r(\delta_n^r)\equiv \mathbb{E}\left[R(\tilde{m}+\delta_n^r,\tilde{m}\mid \tau)\right],
\end{equation}
wherein the expectation is taken with respect to $\tilde{m}$ and $\tau$ according to the joint distribution $f(\tau, \tilde{m})$ given by \eqref{eq:ftautildem}. $\mathcal{R}^r(\delta_n^r)$ can be decomposed into three distinct components corresponding to the different error types shown in \eqref{eq:regret-Pi}:
\begin{equation}\label{eq:R^r}
\begin{aligned}
    \mathcal{R}^r(\delta_n^r)
    &=\int_{-\infty}^{-\delta_n^r} \int_{0}^{\infty} \underbrace{ \Pi(\tau \mid \tau) }_\textbf{Type II Error} f(\tau, \tilde{m}) \, d\tau \, d\tilde{m} + \int_{-\delta_n^r}^{\infty} \int_{-\infty}^{0} \underbrace{ -\Pi(\tilde{m} \mid \tau) }_\textbf{Type I Error} f(\tau, \tilde{m}) \, d\tau \, d\tilde{m}\\
    &+\int_{-\delta_n^r}^{\infty} \int_{0}^{\infty} \underbrace{ \Big[ \Pi(\tau \mid \tau) - \Pi(\tilde{m} \mid \tau) \Big]}_\textbf{Operational Regret} f(\tau, \tilde{m}) \, d\tau  \, d\tilde{m}.
\end{aligned}
\end{equation}

\begin{proposition}[Optimal $\delta_n^r$]
\label{prop:rollout_decision_adjustment}
Suppose Assumption \ref{ass_pitauhat} (i), (iii), (iv), and (v) hold hold.
\begin{enumerate}
    \item There exists a unique optimal rollout decision adjustment $\delta_n^r$ satisfying:
        \begin{equation}\label{eq:prop_1}
            \mathbb{E}_{\tau} \Big[ \Pi(\tilde{m} \mid \tau) \mid \tilde{m} = -\delta_n^r \Big] = 0,
        \end{equation}
    and $\delta_n^r$ satisfies:
    \begin{equation}
          \delta_n^r =
\frac{\Pi^{(0,2)}(0|0)}{2\Pi^{(0,1)}(0|0)}\,\tilde v (1+o(1)).
\end{equation}

    \item The sign of $\delta_n^r$ depends on the curvature of the profit function:
        \begin{enumerate}
            \item If $\Pi(\cdot \mid \tau)$ is linear in $\tau$, then $\delta_n^r=0$;
            \item If $\Pi(\cdot \mid \tau)$ is strictly concave in $\tau$, then $\delta_n^r<0$;
            \item If $\Pi(\cdot \mid \tau)$ is strictly convex in $\tau$, then $\delta_n^r>0$.
        \end{enumerate}
    \item The adjustment $\delta_n^r$ converges to zero at a rate of $O(n^{-1})$. Consequently, in cases 2(b) and 2(c), the regret reduction from adjustment for the rollout decision is $\mathcal{R}^r(0)-\mathcal{R}^r(\delta_n^r)= O(n^{-1})$ as $n\to \infty$.
\end{enumerate}
\end{proposition}

If the SNR function is concave in the treatment effect, as shown in Figure \ref{fig:new_Pi_newsvendor} under the example of inventory management under demand uncertainty, the decision maker suffers disproportionately more from large negative outcomes (losses) than they gain from equivalent large positive outcomes. To compensate for this asymmetry and mitigate ``downside risk,'' the decision maker adopts a \textit{conservative} strategy: they raise the standard of evidence required for implementation ($\delta_n^r < 0$). This implies requiring a posterior confidence level strictly greater than 50\% before taking action.
If the SNR function is convex (as shown in Figure \ref{fig:Pi_wrt_tau_service} in the example of service technology upgrading and capacity planning), the opposite holds (i.e., it encourages an \textit{aggressive} strategy).

\subsection{Adjustment for the Unit Operational Decisions}\label{sec:variable}
In this section, we analyze a regime in which the decision-maker adjusts the estimator solely for the operational decision while leaving the rollout decision unadjusted. Again, we adopt the Bayes decision framework and aim to replace the PTO estimator with the optimal posterior quantile. Specifically, the decision-maker employs $\hat{\tau}^r = \tilde{m}$ for the rollout decision, whereas
\begin{equation}
    \label{eq:tau_n_o}
    \hat{\tau}^o = \tilde{m} + \delta_n^o
\end{equation}
serves as the adjusted estimator for the downstream operational decision and $\delta_n^o$ corresponds to the optimal quantile. Such a quantile-based, additive adjustment rule has two advantages. First, the additive specification yields cleaner asymptotic behavior and avoids scale distortions that may arise under alternative formulations, such as multiplicative adjustments. Specifically, under the concavity of the SNR payoff function $\Pi(\hat{\tau}^o \mid \tau)$ with respect to the estimator $\hat{\tau}^o$, the existence and uniqueness of the optimal adjustment $\delta_n^o$ follow directly. Second, the additive adjustment $\tilde{m} + \delta_n^o$ can isolate the second-order effects of estimation error. A Taylor expansion of the expected regret shows that the leading terms affected by this adjustment are of order $O(n^{-2})$. This structure allows us to exploit a key property of the PTO estimator stated in Lemma~\ref{lemma_tildem_tau}(2), namely that the estimation error $\tilde{m}-\tau$ is independent of $\tilde{m}$, which substantially simplifies the objective function with respect to $\delta_n^o$. Applying Stein’s lemma further reduces the objective to a simple quadratic form in $\delta_n^o$.

Similar to the analysis of the rollout decision, the \textit{prior expected regret} with adjustment $\delta^o_n$ is obtained by integrating over the joint distribution of $(\tau, \tilde{m})$ and is decomposed as follows:
\begin{equation}
\label{eq:R^o}
\begin{aligned}
\mathcal{R}^o(\delta^o_n)&=\int_{-\infty}^{0} \int_{0}^{\infty}  \underbrace{\Pi(\tau\mid\tau) }_\textbf{Type II Error} f(\tau,\tilde{m})\, d\tau  \, d\tilde{m}+  \int_{0}^{\infty} \int_{-\infty}^{0} \underbrace{-\Pi(\tilde{m} + \delta_n^o\mid\tau)}_{\textbf{Type I Error}} f(\tau,\tilde{m}) \, d\tau  \, d\tilde{m}&\\
   & + \int_{0}^{\infty} \int_{0}^{\infty}  \underbrace{\Big[\Pi(\tau\mid\tau) - \Pi(\tilde{m} + \delta_n^o\mid\tau)\Big] }_{\textbf{Operational Regret}} f(\tau,\tilde{m}) \, d\tau \, d\tilde{m}.
   \end{aligned}
\end{equation}

Simplifying $\mathcal{R}^o(\delta^o_n)$ leads us to the observation that
\begin{equation}
\label{eq:last_term}
\mathcal{R}^o(\delta^o_n)=\mathbb{E}\left[ \Pi(\tau\mid\tau) {\mathbb{I}\{\tau>0\}}\right] - \mathbb{E}_{\tau, \tilde{m}} \left[ \Pi(\tilde{m}+\delta^o_n \mid \tau)\cdot \mathbb{I}\{\tilde{m} > 0\}\right].
\end{equation}
Hence, minimizing $\mathcal{R}^o(\delta^o_n)$ is equivalent to maximizing the expected SNR conditional on implementation ($\tilde{m} > 0$). Hereafter, we use the shorthand $\mathbb{E}_{\tau, \tilde{m} > r}[\cdot]$ to denote $\mathbb{E}_{\tau, \tilde{m}}[\cdot \mathbb{I}\{\tilde{m} > r\}]$. Accordingly, $\mathcal{R}^o(\delta^o_n)=\mathbb{E}\left[ \Pi(\tau\mid\tau){\mathbb{I}\{\tau>0\}} \right] - \mathbb{E}_{\tau, \tilde{m} > r} \left[ \Pi(\tilde{m}+\delta^o_n \mid \tau) \right]$. The following proposition presents a closed-form expression for the optimal $\delta_n^o$ and characterizes the resulting improvement in prior expected regret. Detailed derivations are provided in Appendix~\ref{sec:proof_single}.

\begin{proposition}[Optimal $\delta^o_n$]\label{prop:operational_adjustment}
    Suppose Assumption  \ref{ass_pitauhat} (i), (iii), (iv), and (v) holds.
    \begin{enumerate}
        \item There exists a unique optimal $\delta^o_n$ satisfying
        $$\mathbb{E}_{\tau, \tilde{m}>0} \left[ \Pi^{(1,0)}(\tilde{m} + \delta_n^{o}\mid\tau) \right] = 0,$$
         and $\delta_n^o$ satisfies
        \begin{equation}
    \delta^o_n
=
-\frac{\tilde v\,\mathbb{E}_{\tau, \tilde{m}>0}
\left[\Pi^{(1,2)}(\tau\mid\tau)\right]}
{2\,\mathbb{E}_{\tau, \tilde{m}>0}\left[\Pi^{(2,0)}(\tau\mid\tau)\right]}(1+o(1)).
\end{equation}
\item The adjustment $\delta^o_n$ scales as $O(n^{-1})$, and the resulting asymptotic improvement in expected regret, relative to the standard PTO estimator, is given by:
    \begin{align}
\mathcal{R}^o(0)-\mathcal{R}^o(\delta^o_n) =-\frac{1}{8}
\frac{
\left\{\tilde v\,
\mathbb{E}_{\tau, \tilde{m}>0}
\left[\Pi^{(1,2)}(\tau\mid\tau)\right]\right\}^2}
{\mathbb{E}_{\tau, \tilde{m}>0}\left[\Pi^{(2,0)}(\tau\mid\tau)\right]}(1+o(1))
>0.
    \end{align}
This improvement in expected regret scales as $O(n^{-1})$ as $n \to \infty$.
    \end{enumerate}
\end{proposition}
 From Proposition \ref{prop:operational_adjustment}, we directly obtain the following corollaries about the necessity and direction of the adjustment. We discuss the implications immediately after.

\begin{corollary}[Necessity of Adjustment for the Operational Decisions]
\label{corr:variable_adjustment}
If the SNR $\Pi(\hat{\tau}^o \mid \tau)$ exhibits \textit{Non-degenerate Curvature Interaction}, that is $\mathbb{E}_{\tau, \tilde{m}>0}\left[\Pi^{(1,2)}(\tau\mid\tau)\right] \neq 0$, and the estimator is subject to finite-sample noise, then the optimal adjustment $\delta^o_n$ is strictly non-zero, and the magnitude of adjustment is the largest of order $O(n^{-1})$.

\end{corollary}

\begin{corollary}[Direction of Adjustment for the Operational Decisions]\label{corr:variable_adjustment_direction} If the SNR $\Pi(\hat{\tau}^o \mid \tau)$ exhibits \textit{Non-degenerate Curvature Interaction}, that is $\mathbb{E}_{\tau, \tilde{m}>0}\left[\Pi^{(1,2)}(\tau\mid\tau)\right] \neq 0$, and the estimator is subject to finite-sample noise, the sign of the adjustment is the same as the sign of the expected cross-curvature:$$    \operatorname{sgn}(\delta^o_n) = \operatorname{sgn}\left(\mathbb{E}_{\tau, \tilde{m}>0}\left[\Pi^{(1,2)}(\tau\mid\tau)\right]\right).$$\end{corollary}

\paragraph{Economic intuition.}
The second derivative $\Pi^{(2,0)}(\tau \mid \tau)$ captures the local curvature of SNR with respect to the operational decision at the oracle position and therefore governs the local cost of miscalibration when the effect is known. In contrast, the third-order mixed partial derivative $\Pi^{(1,2)}(\tau \mid \tau)$ captures a ``2D'' skewness in SNR. In particular, evaluated at $\hat{\tau}^o=\tau$, it measures how the curvature of $\Pi(\hat{\tau}^o \mid \tau)$ with respect to the true state $\tau$ varies with the implemented decision $\hat{\tau}^o$. Hence, it reflects how under- and over-estimation generate asymmetric losses. Suppose, for example, that $\Pi(\hat{\tau}^o \mid \tau)$ is concave in $\tau$, so that uncertainty in the true state induces a Jensen penalty that increases with the magnitude of curvature. The cross derivative $\Pi^{(1,2)}(\hat\tau^o\mid \tau)$ measures how this curvature (and hence the magnitude of the uncertainty penalty) varies with the adjusted estimator. In particular, $\Pi^{(1,2)}(\tau \mid \tau)>0$ implies that increasing $\hat{\tau}^o$ makes $\Pi$ locally less concave in $\tau$, thereby reducing the cost of estimation error and favoring an upward adjustment. Conversely, $\Pi^{(1,2)}(\tau \mid \tau)<0$ favors a downward adjustment.

\section{Dual Adjustments}
\label{sec:double}
In this section, we consider simultaneously and respectively determining the adjustments to the rollout decision and the unit operational decisions. We first present the formulation and establish theoretical results for the optimal adjustments, and then compare them with results in Section~\ref{sec:single_adjustment}. Finally, we propose an algorithm to compute the adjustments.

\subsection{Formulation and Theoretical Results}
Under dual adjustments, the adjusted estimators for the rollout and unit operational decisions are given respectively by
\begin{equation}\label{eq:double_tau_n_r}
    \hat{\tau}^j = \tilde{m} + \delta_n^{j+},
\end{equation}
wherein $j\in \{r,o\}$. Then, the \textit{prior expected regret} can be written as $ \mathcal{R}(\delta_n^{r+},\delta_n^{o+})=  $
\begin{equation}
\label{eq:R^D}
    \begin{aligned}
&\int_{-\infty}^{-\delta_n^{r+}} \int_{0}^{\infty} \underbrace{ \Pi(\tau\mid\tau) }_\textbf{Type II Error} f(\tau,\tilde{m}) \, d\tau  \, d\tilde{m} + \int_{-\delta_n^{r+}}^{\infty} \int_{-\infty}^{0} \underbrace{-\Pi(\tilde{m}+\delta_n^{o+}\mid\tau)}_{\textbf{Type I Error}}  f(\tau,\tilde{m}) \, d\tau \, d\tilde{m} &\\
  & + \int_{-\delta_n^{r+}}^{\infty} \int_{0}^{\infty}  \underbrace{\Big[\Pi(\tau\mid\tau) - \Pi(\tilde{m}+\delta_n^{o+}\mid\tau)\Big]}_{\textbf{Operational Regret}}f(\tau,\tilde{m}) \, d\tau \, d\tilde{m}.
\end{aligned}
\end{equation}
Observe that the dual-adjustment approach generalizes both previous cases: setting $\delta_n^{r+} = 0$ recovers the single adjustment for the operational decision \eqref{eq:R^o}, while setting $\delta_n^{o+} = 0$ recovers the single adjustment for rollout decision \eqref{eq:R^r}. Then we derive the following theoretical results.
\begin{proposition}[Dual Adjustments]
\label{prop:double}
Suppose Assumption \ref{ass_pitauhat} holds, under the dual adjustments strategy:
\begin{enumerate}
    \item The optimal adjustments $(\delta_n^{r+},\delta_n^{o+})$ satisfy the following conditions:
\begin{align}
\label{eq:eta^D}
\mathbb{E}_{\tau} \left[ \Pi(\tilde{m} + \delta_n^{o+}| \tau) \middle| \tilde{m}=-\delta_n^{r+} \right] &= 0, \\
\label{eq:tau_K^D}
\mathbb{E}_{\tau, \tilde{m}>-\delta_n^{r+}} \left[ \Pi^{(1,0)}(\tilde{m} + \delta_n^{o+}\mid\tau) \right] &= 0. \end{align}
The optimal $\delta_n^{o+}$ can be approximated as a function of $\delta_n^{r+}$:
\begin{equation}
\label{eq:lambda(tau_k)}
\delta_n^{o+}(\delta_n^{r+})=\frac{\tilde v\,\mathbb{E}_{\tau, \tilde{m}>-\delta_n^{r+}}
\left[\Pi^{(1,2)}(\tau\mid\tau)\right]}
{2\,\mathbb{E}_{\tau, \tilde{m}>-\delta_n^{r+}}\left[\Pi^{(2,0)}(\tau\mid\tau)\right]}(1+o(1)).
\end{equation}
    \item The adjustments $\delta_n^{j+},j=\{r,o\}$, converge to zero at rate $O(n^{-1})$ whenever they are nonzero. Consequently, the regret reduction induced by the dual adjustments satisfies $\mathcal{R}(0,0)-\mathcal{R}(\delta_n^{r+},\delta_n^{o+})= O(n^{-1})$ as $n\to \infty$, provided that at least one of $\delta_n^{j+},j=\{r,o\}$ is nonzero.
\end{enumerate}

\end{proposition}

From \eqref{eq:eta^D} and Assumption~\ref{ass_pitauhat}(iii), the following result follows directly from Jensen’s inequality.
\begin{corollary}
\label{corr:concave_taukD}
    If $\Pi(\hat{\tau}^o \mid \tau)$ is strictly concave in the true effect $\tau$, then $\delta_n^{r+}<0$.
\end{corollary}

We next identify a condition under which the optimal adjustment for the operational decisions is independent of the adjustment for rollout decision, that is, $\delta_n^{o+}=\delta^o_n$. In this case, the dual-adjustment problem decouples into a \textit{sequential optimization process}. The decision maker can first determine the optimal adjustment $\delta_n^{o+}=\delta^o_n$ (via Proposition \ref{prop:operational_adjustment}), and then solve for the optimal adjustment for rollout decision $\delta_n^{r+}$ by incorporating this $\delta_n^{o+}$ (via Proposition \ref{prop:double}).

\begin{corollary}\label{corr:independence}
The optimal adjustment for unit operational decisions is independent of the adjustment for rollout decision (i.e., $\delta_n^{o+} = \delta^o_n$) if both $\Pi^{(1,2)}(\tau\mid\tau)$ and $\Pi^{(2,0)}(\tau\mid\tau)$ are non-zero, and their ratio $$\frac{\Pi^{(1,2)}(\tau\mid\tau)}{\Pi^{(2,0)}(\tau\mid\tau)}\equiv \text{constant},$$ for all $\tau$. This condition is satisfied when $\Pi(\hat{\tau}^o \mid \tau)$ is \textbf{additively separable with a difference kernel}:
\begin{align}
\label{eq:separable}
\Pi(\hat{\tau}^o \mid \tau) = h(\tau) + g(\hat{\tau}^o-\tau),
\end{align}
where $h(\cdot)$ and $g(\cdot)$ are arbitrary functions.
\end{corollary}

In the inventory management example under demand uncertainty (Section~\ref{sec:newsvendor_example}), $\Pi(\hat{\tau}^o \mid \tau)$ satisfies the separable structure in \eqref{eq:separable}, and both $\Pi^{(1,2)}(\tau \mid \tau)$ and $\Pi^{(2,0)}(\tau \mid \tau)$ are nonzero constants. Consequently, we obtain $\delta_n^{o+}=\delta_n^o$.
In the service operations example with capacity planning (Section~\ref{sec:service_time_example}), both $\Pi^{(1,2)}(\tau \mid \tau)$ and $\Pi^{(2,0)}(\tau \mid \tau)$ depend on $\tau$. Nevertheless, their ratio remains constant because $\Pi^{(1,2)}(\tau \mid \tau)=\Pi^{(2,0)}(\tau \mid \tau)$ for all $\tau$. Therefore, $\delta_n^{o+}=\delta_n^o$ again holds.


\subsection{Interaction Between the Two Adjustments}
In Section~\ref{sec:single_adjustment}, we consider single-adjustment approaches
(i.e., adjustment for the rollout decision or operational decision only). To assess the effects of allowing both
adjustments, we compare the magnitude of the adjustment for rollout decision under the dual adjustments strategy, $\delta_n^{r+}$, with that under the single-adjustment strategy, $\delta_n^r$.

\begin{definition}[Substitution and Complementarity of Adjustments]
\label{def:substitution_complementarity}
The adjustments for rollout and operational decisions are defined as \textit{substitutes} if $|\delta_n^{r+}| < |\delta_n^r|$, and as \textit{complements} if $|\delta_n^{r+}| > |\delta_n^r|$.
\end{definition}

The intuition is as follows. If $|\delta_n^{r+}| < |\delta_n^r|$ (the case of \textit{substitutes}), the operational adjustment absorbs part of the corrective burden, thereby reducing the magnitude of the rollout adjustment under the dual adjustments strategy. In contrast, if $|\delta_n^{r+}| > |\delta_n^r|$ (the case of \textit{complements}), the operational adjustment reinforces the rollout adjustment, resulting in a larger magnitude of the rollout adjustment when both instruments are employed simultaneously. Proposition~\ref{prop:complementary} below establishes a condition for complementarity.

\begin{proposition}[Conditions for Complementarity]\label{prop:complementary}
The rollout adjustment is weakly larger in magnitude under dual adjustments than under single adjustment (i.e., $|\delta_n^{r+}| \geq |\delta_n^r|$) if the following conditions hold:

\medskip
\noindent\textbf{Case I:} $\delta_n^{r+} \geq \delta_n^r >0$ holds if $\Pi(\hat{\tau}^o \mid \tau)$ is convex in $\tau$, and one of the following condition is satisfied:
\begin{enumerate}
    \item[(a)] $\Pi(\hat{\tau}^o \mid \tau) = h(\tau) + g(\hat{\tau}^o - \tau)$ is \textit{additively separable}, as defined in \eqref{eq:separable}.

    \item[(b)] $\Pi^{(1,2)}(\tau\mid\tau)=\eta\Pi^{(2,0)}(\tau\mid\tau)$ for all $\tau$, wherein $\eta$ is a constant.
    \item[(c)] $\Pi^{(1,2)}(\tau\mid\tau)=\eta(\tau) \cdot \Pi^{(2,0)}(\tau\mid\tau)$ for all $\tau$, wherein $\eta(\tau)$ is a function of constant sign, and $|\eta(-\delta_n^r)|>\bar{\eta}/2$, wherein    $$\bar{\eta}\equiv\frac{\mathbb{E}_{\tau, \tilde{m}>-\delta_n^{r+}}\left[\Pi^{(1,2)}(\tau\mid\tau)\right]}{\mathbb{E}_{\tau, \tilde{m}>-\delta_n^{r+}}\left[|\Pi^{(2,0)}(\tau\mid\tau)|\right]}.$$
    \end{enumerate}

\medskip
\noindent\textbf{Case II:} $\delta_n^{r+} \leq \delta_n^r < 0$ holds if $\Pi(\hat{\tau}^o \mid \tau)$ is concave in $\tau$ and $\Pi^{(1,2)}(-\delta_n^r|-\delta_n^r)\leq0$.

\medskip
\noindent Furthermore, the strict inequalities $\delta_n^{r+} > \delta_n^r$ (Case I) and $\delta_n^{r+} < \delta_n^r$ (Case II) hold whenever $\delta_n^{o+} \neq 0$.
\end{proposition}

In Proposition \ref{prop:complementary} Case I, condition (a) is a special case of condition (b), while condition (b) is a special case of condition (c). The example of service operations with capacity planning in Section \ref{sec:service_time_example} satisfies condition (b), resulting in $\delta_n^{r+} > \delta_n^r > 0$, as shown in Figure \ref{fig:service_combined}.

We now derive sufficient conditions under which two adjustment serves as substitute. From Corollary~\ref{corr:concave_taukD} and Proposition~\ref{prop:complementary}, the following result follows immediately.

\begin{corollary}[Conditions for Substitution]\label{prop:larger_thresholds}
The rollout adjustment is weakly smaller in magnitude under dual adjustments than under single adjustment (i.e., $|\delta_n^r| \ge |\delta_n^{r+}|$, where $\delta_n^r, \delta_n^{r+} < 0$) if the following conditions hold:
$\Pi(\hat{\tau}^o \mid \tau)$ is concave in $\tau$, and one of the conditions in Proposition \ref{prop:complementary} Case I is satisfied. Furthermore, the strict inequalities $\delta_n^r< \delta_n^{r+}<0$ hold whenever $\delta_n^{o+} \neq 0$.
\end{corollary}

\subsection{Algorithm for Dual Adjustments and Convergence Analysis}
\begin{algorithm}[htbp]
\caption{Alternating Iteration for Dual Adjustments}
\label{alg:alternating}
\begin{algorithmic}[1]

\Require Initial guesses $r^{(0)} = 0$, $o^{(0)} = 0$; tolerance $t > 0$
\Ensure Optimal adjustments $\delta_n^{r+}$, $\delta_n^{o+}$

\For{$k = 0,1,2,\ldots$}
  \State \textbf{Step A:} Given $r^{(k)}$, compute $o^{(k+1)}$ such that
  \State \hspace{1em}$\mathbb{E}_{\tau,\, \tilde m > -r^{(k)}}
  \!\left[\Pi^{(1,0)}(\tilde m + o^{(k+1)} \mid \tau)\right] = 0$

  \State \textbf{Step B:} Given $o^{(k+1)}$, compute $r^{(k+1)}$ such that
  \State \hspace{1em}$\mathbb{E}_{\tau}
  \!\left[\Pi(\tilde m + o^{(k+1)} \mid \tau)\,\middle|\, \tilde m = -r^{(k+1)}\right] = 0$

  \If{$|r^{(k+1)}-r^{(k)}| + |o^{(k+1)}-o^{(k)}| < t$}
    \State \textbf{Terminate}
    \State \textbf{break}
  \EndIf
\EndFor

\State \Return $(\delta_n^{r+}, \delta_n^{o+}) = (r^{(k+1)}, o^{(k+1)})$

\end{algorithmic}
\end{algorithm}
We now establish the convergence of Algorithm~\ref{alg:alternating}.

\begin{theorem}[Convergence of Algorithm~\ref{alg:alternating}]
\label{thm:global_convergence}
Suppose Assumption~\ref{ass_pitauhat} (i), (ii), (iv), and (v) hold.
Assume further that there exist constants $\kappa_o>0$, $\kappa_r>0$, $\rho>0$, and $\bar n>0$ such that for all $n>\bar n$ and all $|r|\le \rho/n$, $|o|\le \rho/n$ the following conditions hold:

\begin{enumerate}
\item[(C1)] \textbf{Local curvature of the operational mapping:}
\[
-\mathbb{E}_{\tau}\!\left[
\Pi^{(2,0)}(\tilde m+o \mid \tau)
\;\middle|\;
\tilde m > -r
\right]
\;\ge\;
n\,\kappa_o .
\]

\item[(C2)] \textbf{Local nondegeneracy of the rollout mapping:}
\[
\left|
\mathbb{E}_{\tau}\!\left[
\Pi^{(1,0)}(-r+o \mid \tau)
+
\Pi^{(0,1)}(-r+o \mid \tau)
\;\middle|\;
\tilde m = -r
\right]
\right|
\;\ge\;
n\,\kappa_r .
\]
\end{enumerate}
Then there exists $\bar{n} > 0$ such that for all $n > \bar{n}$:
\begin{enumerate}
    \item[(i)] There exists a unique solution $(\delta_n^{r+}, \delta_n^{o+})$ to the system \eqref{eq:eta^D}--\eqref{eq:tau_K^D}.
    \item[(ii)] Algorithm~\ref{alg:alternating} converges to $(\delta_n^{r+}, \delta_n^{o+})$ at a linear rate.
\end{enumerate}
\end{theorem}


\section{Benchmarking against PTO and the Bayes Rule}
\label{sec:benchmark}
Because PATRO sits between “pure plug-in” decision making and fully Bayesian decision making, it is natural to benchmark it against two canonical reference rules:
\begin{enumerate}
    \item \textbf{Predict-Then-Optimize (PTO).} PTO plugs the posterior mean (or sample mean difference) into both the rollout rule and the downstream optimizer without any correction for finite-sample uncertainty, corresponding to PATRO with zero adjustments (i.e., using the posterior median/mean quantile). Therefore, PTO is dominated by PATRO. In Section~\ref{sec:numerical}, we show PATRO improves the prior expected regret against PTO.
    \item \textbf{The Bayes-optimal decision rule.}
The Bayes rule chooses the rollout and downstream decisions to maximize
posterior expected payoff conditional on the observed data.
Unlike plug-in approaches, it allows the decision to depend on the entire
posterior distribution rather than on a single adjusted point estimate.
Accordingly, it represents the strongest achievable benchmark within our
Bayesian framework and provides a natural lower bound on expected regret.
Nevertheless, we show that PATRO can coincide with the Bayes rule for certain
nontrivial classes of problems, or exhibit negligible differences in terms of
prior expected regret. For instance, in newsvendor-style problems and pricing
problems with log-linear demand (see Section~\ref{sec:newsvendor_example} and \ref{sec:pricing_example}), the percentage
difference in prior expected regret between PATRO and the Bayes rule is zero,
as demonstrated both numerically and theoretically (Theorem~\ref{thm:benchmark}).
For service capacity management problems (see Section~\ref{sec:service_time_example}),
the percentage difference in prior expected regret is at most on the order of
$10^{-3}$ (e.g., $4\times 10^{-3}\%$), under the same parameter choices as in
Table~\ref{tab:improvement_rates}.

\end{enumerate}
We formalize the relationships between PATRO and the Bayes rule in the following theorem.


\begin{theorem}[Bayes dominance and equivalence with PATRO]
\label{thm:benchmark}

Let $\mathbf{S}=\mathbf{s}\in \mathcal{S}$ be the realized experiment data. Let the two-stage decisions be $\mathbbm a=(D,\mathbf u)\in \mathcal{A}$ with payoff $\Gamma(\mathbbm a,\tau)$ as defined in \eqref{eq:Gamma_function}. Let $\mathbbm a^\#(\tau)=(D^{\#}(\tau),\mathbf u^{\#}(\tau))$ be the oracle decision rule, and $\mathbbm a^*:\mathcal{S}\to \mathcal{A}$ be the Bayes optimal rule such that
\begin{equation}
 \mathbbm a^*(\mathbf{s})\equiv (D^*(\mathbf{s}),\mathbf u^*(\mathbf{s}))\in\arg\max_{\mathbbm a\in \mathcal A}\ \mathbb E[\Gamma(\mathbbm a,\tau)\mid \mathbf{s}].
\end{equation}
Let PATRO restrict attention to constant-posterior-quantile-induced oracle decisions:
\begin{equation}
\mathbb Q=\Big\{\mathbbm v_{\mathbf q}(\mathbf s)=\big(D^{\#}F^{-1}_{\tau\mid \mathbf S=\mathbf s}(q^r)),\mathbf u^{\#}(F^{-1}_{\tau\mid \mathbf S=\mathbf s}(q^o))\big):
\mathbf q=(q^r,q^o)\in(0,1)^2\text{ and }\mathbf s \in \mathcal S\Big\},
\end{equation}
and the PATRO decision rule is $\mathbbm v^*: \mathcal S\rightarrow \mathcal A$ such that
\begin{equation}
\mathbbm v^*\in\arg\max_{\mathbbm v\in \mathbb Q}\ \mathbb E_{\tau,\mathbf S}\big[\Gamma(\mathbbm v(\mathbf S),\tau)\big].
\end{equation}

\begin{enumerate}

\item \textbf{PATRO is weakly dominated by the Bayes rule}, that is
\begin{equation}\label{eq:Bayes-dominance}
    \mathbb E_{\tau,\mathbf S}\big[\Gamma(\mathbbm a^*(\mathbf S),\tau)\big]
\ \ge\
\mathbb E_{\tau,\mathbf S}\big[\Gamma(\mathbbm v^*(\mathbf S),\tau)\big].
\end{equation}

\item \textbf{PATRO is equivalent to Bayes rule} (i.e., equality is achieved in \eqref{eq:Bayes-dominance}) if Assumption \ref{ass_pitauhat} (i)-(v) holds, $\mathrm{Img}[\mathbf u^*]\subseteq\mathrm{Img}[\mathbf u^{\#}]$, and either of the following conditions holds:
\begin{enumerate}
    \item There exists a constant $\bar \delta$ such that, for any square-integrable random variable $Y$ with a mean $\mu_Y$ and any $x\in\mathbb R$, we have
$
\mathbb E[\Pi(x\mid Y)]\le \mathbb E[\Pi(\mu_Y+\bar \delta\mid Y)]
$.
\item $\Pi(\cdot\mid\cdot)$ is additively separable with a difference kernel as defined in \eqref{eq:separable}.
\end{enumerate}
\end{enumerate}
\end{theorem}

When the equivalence conditions hold, the Bayes-optimal operational decisions align with the plug-in decisions based on a posterior quantile, and the remaining improvement from full Bayes optimization (if any) is driven solely by the rollout decision, for which the optimal adjustment does not rely on the posterior mean.

In addition to its strong or even perfect performance, PATRO offers advantages in transparency and implementability against the Bayes-optimal decision rule. Because the optimization is conducted ex ante, the implementation of PATRO requires no additional ex post computation beyond the effect estimation step.

\section{Numerical Examples}\label{sec:numerical}
In this section, we present three examples to verify our theoretical results and illustrate the diversity of optimal adjustment strategies across different economic settings that differ in the structure of the underlying SNR function $\Pi(\hat \tau^o\mid\tau)$.

The first example is a \textit{newsvendor-style demand and inventory management problem}. The decision maker experiments to assess whether the treatment can improve the expected demand, and then decides whether to roll out the treatment and how to re-optimize the ordering decision. We show that, for the rollout decision, the decision maker should adopt a more aggressive implementation strategy (i.e., $\delta_n^r > 0$), and that the rollout and operational adjustments act as substitutes.

The second example is a \textit{technology upgrading and capacity planning problem} for a service provider (such as a restaurant chain). The decision maker experiments to evaluate whether the new technology can improve the long-run service rate, and then determines whether to roll out the upgrade and how to re-optimize the service capacity in each unit. We get the opposite results: the rollout adjustment should be more conservative (i.e., $\delta_n^r < 0$), and the two adjustments act as complements.

The third example is a \textit{promotion and pricing problem}, wherein the decision maker experiments to evaluate whether the promotion can improve the mean potential demand and decides whether to roll out promotion and how to re-optimize downstream pricing. We consider a general demand function that subsumes both linear and log-linear demand as special cases. In both cases, the two adjustments are either non-existent or independent.

Taken together, these examples demonstrate that the optimal adjustment strategy is highly context-dependent and is determined by the specific structural form of the SNR (i.e., $\Pi(\hat{\tau}^o \mid \tau)$). Nevertheless, most examples exhibit a notable reduction in regret achieved through estimator adjustments.
Table~\ref{tab:improvement_rates} summarizes the improvement rates in regret, defined as
\begin{equation}
\label{eq:IR_PATRO}
[\mathcal{R}(0,0)-\mathcal{R}(\delta_n^{r+},\delta_n^{r+})]/\mathcal{R}(0,0)
\end{equation}
where $\mathcal{R}(\cdot)$ is defined in \eqref{eq:R^D}, $\mathcal{R}(\delta_n^{r+},\delta_n^{r+})$ represents PATRO, and $\mathcal{R}(0,0)$ represents PTO.

\begin{table}[ht]
\centering
\small
\begin{threeparttable}
\caption{Regret Improvement Rate (\%) under Different Examples}
\label{tab:improvement_rates}
\renewcommand{\arraystretch}{1.1} 
\begin{tabular}{@{}llccccc@{}} 
\toprule
& & \multicolumn{5}{c}{\textbf{Sample Size ($n$)}} \\
\cmidrule(l){3-7}
\textbf{Type} & \textbf{Examples \& Parameters} & 10 & 30 & 50 & 70 & 90 \\
\midrule
\multirow{3}{*}{\begin{tabular}[c]{@{}l@{}}Demand \& \\Inventory \\ Management\end{tabular}}
& $CR=0.11, v_0=1, \sigma_\varepsilon=1$ & 4.5712 & 2.0392 & 1.3156 & 0.9713 & 0.7699 \\
& $CR=0.11, v_0=2, \sigma_\varepsilon=2$ & 3.6956 & 1.7907 & 1.1888 & 0.8904 & 0.7119 \\
& $CR=0.80, v_0=2, \sigma_\varepsilon=1$ & 4.2093 & 1.7643 & 1.1178 & 0.8182 & 0.6453 \\
\midrule
\multirow{3}{*}{\begin{tabular}[c]{@{}l@{}}Service \\ Capacity \\ Management\end{tabular}}
& $c=0.5, a=1, v_0=1, \sigma_\varepsilon=1$ & 5.1897 & 2.5627 & 1.6962 & 1.2670 & 1.0110 \\
& $c=0.5, a=1, v_0=4, \sigma_\varepsilon=2$ & 8.2418 & 6.7355 & 5.0914 & 4.0461 & 3.3467 \\
& $c=0.5, a=2, v_0=1, \sigma_\varepsilon=2$ & 8.2997 & 5.9306 & 4.6057 & 3.7557 & 3.1674 \\
\midrule
\multirow{3}{*}{\begin{tabular}[c]{@{}l@{}}Pricing with \\ Log-Linear \\ Demand\end{tabular}}
& $a=1, b=1, v_0=2, \sigma_\varepsilon=1$ & 8.4664 & 3.1421 & 1.9294 & 1.3922 & 1.0890 \\
& $a=4, b=1, v_0=5, \sigma_\varepsilon=1$ & 9.1387 & 3.2324 & 1.9633 & 1.4097 & 1.0997 \\
& $a=1, b=1, v_0=5, \sigma_\varepsilon=2$ & 28.8736 & 11.8411 & 7.4401 & 5.4234 & 4.2667 \\
\bottomrule
\end{tabular}
\begin{tablenotes}[para, flushleft]
    \footnotesize
    \textit{Notes.} Parameters common to all examples are $\gamma=1$ and $m_0=0$. Specific settings include $\mu=10, p=10$ for inventory and $p=2$ for service operations.
\end{tablenotes}
\end{threeparttable}
\end{table}

\subsection{Demand and Inventory Management}\label{sec:newsvendor_example}
In this section, we consider a classic newsvendor framework. Let the potential demand in treatment state $D$ outcomes be $X^D=D\cdot \tau+\varepsilon$,
where $\varepsilon\sim \mathcal{N}(\mu, \sigma_\varepsilon^2)$ is the baseline demand with mean $\mu$ and variance $\sigma_\varepsilon^2$ and $\tau$ is the average treatment effect (ATE). The treatment shifts the mean but leaves the variance unchanged.

The firm faces asymmetric costs: $c_u$ per unit of unmet demand (understocking) and $c_o$ per unit of excess inventory (overstocking). The optimal order quantity balances these costs by targeting the critical ratio:
$CR = c_u/(c_u + c_o)$, which represents the service level—the probability of fully satisfying demand. Let $z_{CR} = \Phi^{-1}(CR)$ denote the corresponding standard normal quantile. The optimal order quantity with the treated demand is:
\begin{equation*}
Q(\tau) = \mu + \tau + \sigma_\varepsilon \cdot z_{CR}.
\end{equation*}
In practice, $\tau$ is unknown and must be estimated. Given the adjusted estimate $\hat{\tau}^o$, the firm's order quantity is:
\begin{equation*}
\hat{Q}\equiv Q(\hat{\tau}^o) = \mu + \hat{\tau}^o + \sigma_\varepsilon \cdot z_{CR}.
\end{equation*}
Note that in this example (also for the following example), we assume that all units share homogeneous expected demand ($\mu$), operational costs ($c_u$ and $c_o$), and unit price ($p$). This assumption is made without loss of generality. These parameters can be interpreted as averages across units; in this case, the results remain unchanged.

The expected total payoff conditional on implementation, the estimate $\hat{\tau}^o$, and the true effect $\tau$ is:
\begin{equation}
    \label{eq:newsvendor_profit_raw}
\begin{aligned}
\sum_{i=1}^n G_i\left(Q(\hat \tau^o)\mid\tau\right)
&= n\cdot \mathbb{E}_{X^1}\left[p\cdot X^1 - c_o \cdot [Q(\hat \tau^o) - X^1]^+ - c_u \cdot [X^1 - Q(\hat \tau^o)]^+\right]&\\
&=n\left\{p(\mu + \tau) - (c_u + c_o)\sigma_\varepsilon \phi(z_1) - (c_u + c_o)\sigma_\varepsilon z_1 \Phi(z_1) + c_u \sigma_\varepsilon z_1\right\},&
\end{aligned}
\end{equation}
where $p$ is the unit price, $(\cdot)^+$ denotes the positive part, $z_1$ is the standardized decision error, defined as
\begin{equation*}
z_1 \equiv \frac{Q(\hat \tau^o) - (\mu+\tau)}{\sigma_\varepsilon} = z_{CR} + \frac{\hat\tau^o - \tau}{\sigma_\varepsilon}.
\end{equation*}

In addition, when left untreated, the expected profit is $\tau=\hat\tau^o=0$ in \eqref{eq:newsvendor_profit_raw}, expressed as follows:
\begin{equation}
    \label{eq:G_0_newsvendor}
    G_0 = n\cdot \left[p\mu - (c_u + c_o)\sigma_\varepsilon \phi(z_{CR})\right].
\end{equation}
Hence, after normalizing the implementation cost $c$ to zero, we have the SNR function:
\begin{equation}
\Pi(\hat{\tau}^o \mid \tau) = n\left\{p\tau - (c_u + c_o)\sigma_\varepsilon [\phi(z_1)-\phi(z_{CR})] - (c_u + c_o)\sigma_\varepsilon z_1 \Phi(z_1) + c_u \sigma_\varepsilon z_1\right\}.
\label{eq:newsvendor_SNR}
\end{equation}
where $\phi(\cdot)$ and $\Phi(\cdot)$ denote the standard Normal p.d.f and c.d.f, respectively. This SNR function can be shown to satisfy all assumptions stated in Assumption~\ref{ass_pitauhat} and the conditions of Theorem~\ref{thm:global_convergence}.

\begin{figure*}[htbp]
\centering
\begin{subfigure}{0.95\textwidth}
    \centering   \includegraphics[width=\textwidth]{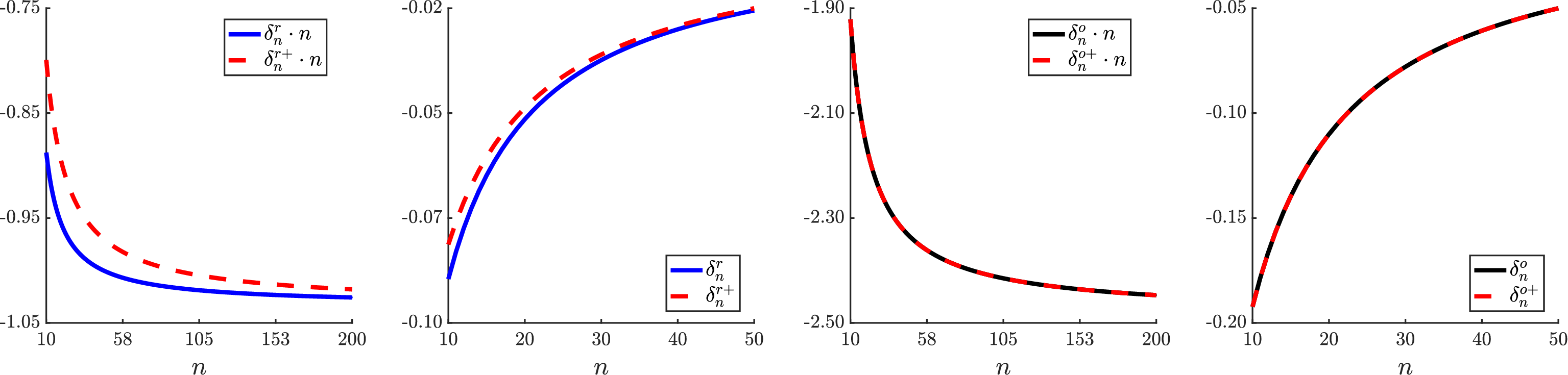}
    \caption{The convergence rate of adjustments}
\label{fig:convergence_newsvendor}
\end{subfigure}
\vspace{2ex}
\begin{subfigure}{0.95\textwidth}
    \centering
    \includegraphics[width=\textwidth]{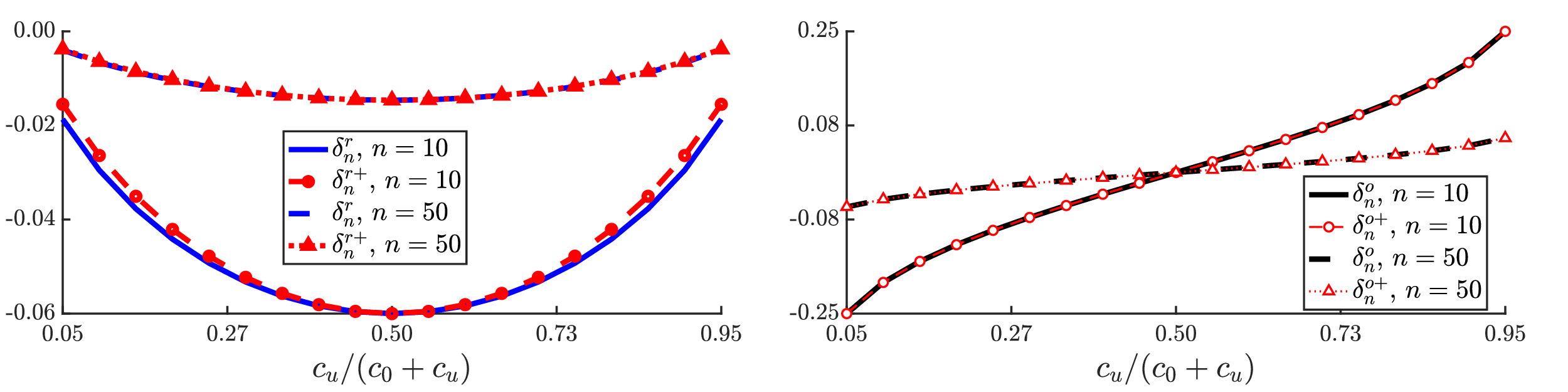}
    \caption{The impact of critical ratio}
    \label{fig:critical_ratio}
\end{subfigure}

\vspace{2ex}

\begin{subfigure}[b]{0.45\textwidth}
        \centering
        \includegraphics[width=\textwidth]{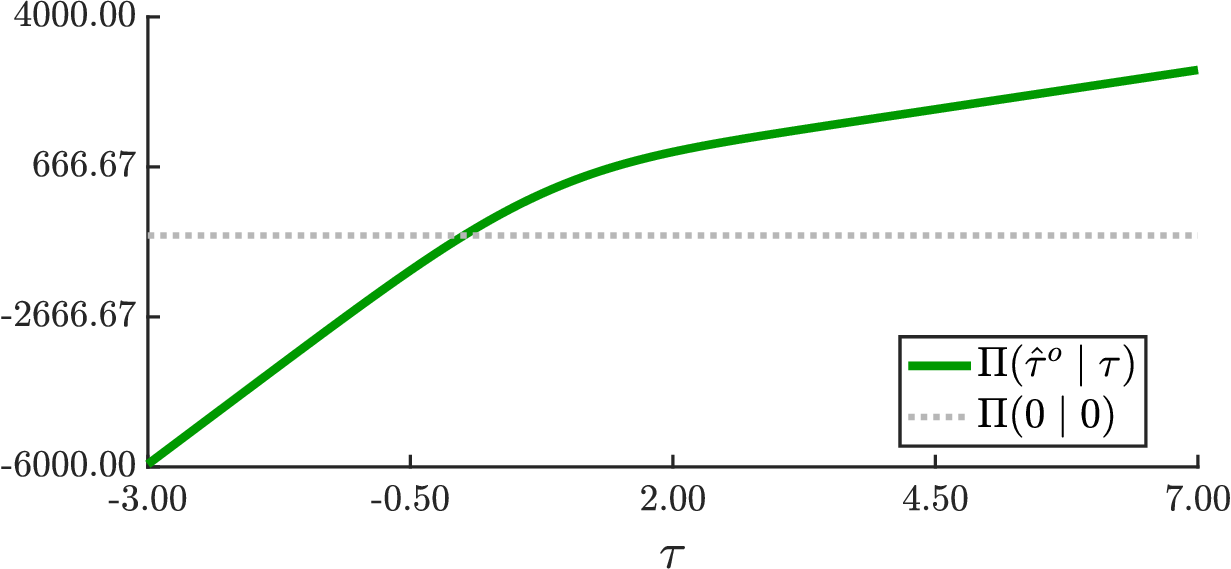}
        \caption{$\Pi(\hat{\tau}^o\mid\tau)$ given $\hat{\tau}^o=2,n=50$}
        \label{fig:new_Pi_newsvendor}
    \end{subfigure}%
    \hspace{0.05\textwidth}
\begin{subfigure}{0.45\textwidth}
    \centering
    \includegraphics[width=\textwidth]{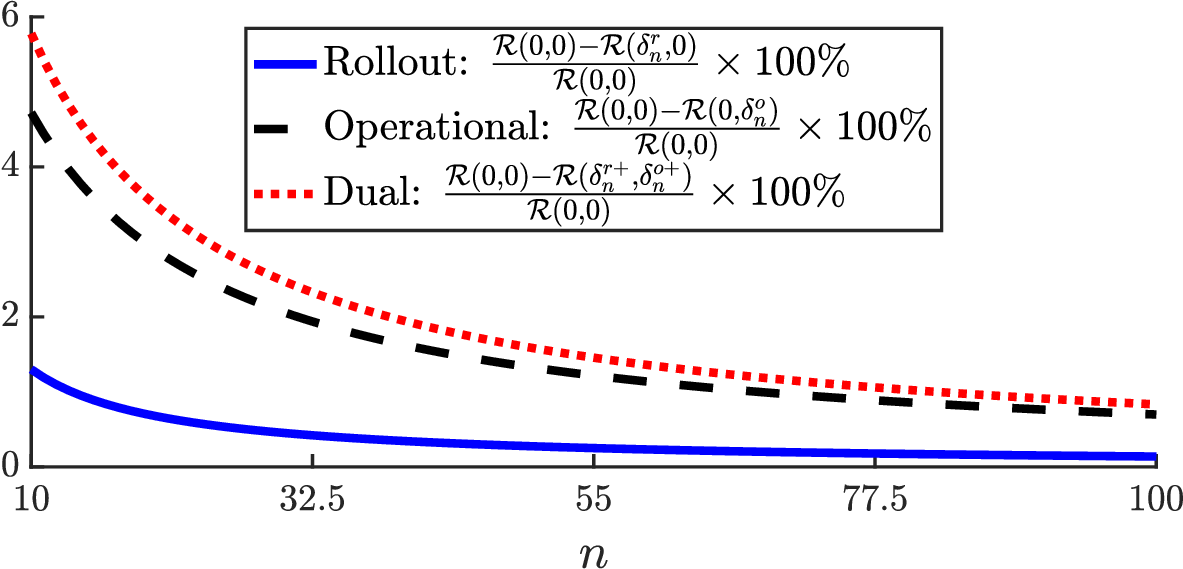}
    \caption{Regret improvement rate}
    \label{fig:regret_improvement_newsvendor}
\end{subfigure}
\caption{Inventory Management under Demand Uncertainty}
\label{fig:combined_analysis}

\vspace{1ex}

\parbox{\textwidth}{
    \footnotesize
    \textit{Note:}
    Parameters: $\mu=10, p=10, m_0=0, v_0=2, \sigma_\varepsilon=1, \gamma=1$.
    For plots (a), (c) and (d): $c_u=3, c_o=25$.
    For plot (b): $c_u + c_o = 10$ with $CR = c_u/(c_u+c_o)$ varying from 0.05 to 0.95, and $n \in \{10, 50\}$.
}
\end{figure*}

\vspace{-1em}
\paragraph{Adjustment for rollout decision} Computing the partial derivatives with respect to $\tau$, we obtain
\begin{align*}
    \Pi^{(0,1)}(\hat{\tau}^o\mid\tau) = &n[p+(c_u+c_o)\Phi(z_1)-c_u], \,\ \forall \hat{\tau}^o&\\
    \Pi^{(0,2)}(\hat{\tau}^o\mid\tau) =& -n[(c_u+c_o)\phi(z_1)/\sigma_\varepsilon]<0, \,\ \forall \hat{\tau}^o&
\end{align*}
The strict negativity of the second derivative establishes that $\Pi(\cdot \mid \tau)$ is concave in $\tau$ (see Figure~\ref{fig:new_Pi_newsvendor}). By Proposition~\ref{prop:rollout_decision_adjustment}, this implies a negative adjustment for rollout decision:    \begin{equation}
\delta_n^r =
\frac{\Pi^{(0,2)}(0|0)}{2\Pi^{(0,1)}(0|0)}\,\tilde v (1+o(1))\approx -\frac{(c_u+c_o)\phi(z_{CR})\tilde v}{2p\sigma_\epsilon}<0,
\end{equation} which is also verified numerically in Figure~\ref{fig:convergence_newsvendor}.
The concavity reflects diminishing marginal returns with respect to the true treatment effect. Once inventory $\hat{Q}$ is committed, the payoff structure becomes asymmetric:
\begin{itemize}
\item \textbf{Downside exposure:} If $\tau < \hat{\tau}^o$, actual demand falls below expectations. Excess inventory incurs overstocking costs proportional to the gap, and profit declines steeply.
\item \textbf{Capped upside:} If $\tau > \hat{\tau}^o$, demand exceeds expectations. However, sales are constrained by available inventory $\hat{Q}$, so profit gains flatten as stockouts become more likely.
\end{itemize}

\vspace{-1em}
\paragraph{Adjustment for the operational decision}
In the example of demand and inventory management, we have
    $\Pi^{(2,0)}(\tau\mid\tau) = -n[(c_u+c_o)\phi(z_{CR})/\sigma_\varepsilon]$ and $
    \Pi^{(1,2)}(\tau\mid\tau) = n[(c_u+c_o) z_{CR} \phi(z_{CR})/\sigma_\varepsilon^2]$.
Hence,
\begin{equation*}
    \delta^o_n
=
-\frac{\tilde v\,\mathbb{E}_{\tau, \tilde{m}>0}
\left[\Pi^{(1,2)}(\tau\mid\tau)\right]}
{2\,\mathbb{E}_{\tau, \tilde{m}>0}\left[\Pi^{(2,0)}(\tau\mid\tau)\right]}=\frac{\tilde{v}z_{CR}}{2\sigma_\varepsilon}.
    \end{equation*}
wherein $\tilde{v}
=
\left[
v^{-1}
+
\gamma n/(\sigma_\varepsilon^2(1+\gamma)^2)
\right]^{-1}$ defined in \eqref{eq:vtilde_mtilde}.

When $z_{CR}>0$
(equivalently, $CR>0.5$), the understocking cost exceeds the overstocking cost ($c_u>c_o$). In this case, $\delta_n^o > 0$: the decision maker should order more aggressively to guard against stockouts. Conversely, when $z_{CR}<0$ (equivalently, $CR<0.5$), overstocking is more costly ($c_o > c_u$), and $\delta_n^o < 0$: the decision maker should order conservatively. Figure~\ref{fig:critical_ratio} confirms this relationship. Interestingly, the size of rollout adjustment $\delta_n^r$ is the largest when $CR=0.5$. This is because in this case the adjustment for $Q$ is not necessary due to the symmetry, resulting in the highest levels of Type I Error and operational regret. To minimize the overall regret, the decision maker needs to impose the highest bar for approving the rollout.

\vspace{-1em}
\paragraph{Dual adjustments.} Since $\Pi(\hat{\tau}^o \mid \tau)$ is concave in $\tau$ and the SNR function in \eqref{eq:newsvendor_SNR} satisfies condition (a) (separability), Proposition \ref{prop:complementary} implies that $\delta_n^r < \delta_n^{r+} < 0$, which is verified in Figure \ref{fig:convergence_newsvendor}.

\subsection{Service Technology Upgrading and Capacity Planning}\label{sec:service_time_example}
In this example, the decision maker manages a service system under heavy traffic and considers expediting the service process. He first evaluates whether the treatment can improve the long-run average service rate. The baseline service time per task follows a log-normal distribution \citep{brown2005statistical}:
\[
\log T^0 = a + \varepsilon,
\]
where $a$ is the nominal log-service-time parameter and $\varepsilon \sim \mathcal{N}(0, \sigma_\varepsilon^2)$ captures random variability. A new technology (e.g., automation or better communication) reduces the log-service time by $\tau$, yielding treated service time:
\[
\log T^1 = a - \tau + \varepsilon.
\]
The decision maker estimates $\tau$ from experimental data, using $\hat\tau^o$ after adjustment, and provides $M(\hat\tau^o)$ servers to maximize long-run average profit.

Let $\{T_i^1\}_{i \geq 1}$ be i.i.d.\ service times under the treatment. Since $T^1 = \exp(a - \tau + \varepsilon)$ is log-normal with mean $\mathbb{E}[T^1] = \exp(a - \tau + \sigma_\varepsilon^2/2)$, the elementary renewal theorem gives the long-run service completion rate:
\[
\lambda(\tau) := \lim_{t \to \infty} \frac{N(t)}{t} = \frac{1}{\mathbb{E}[T^1]} = \exp\left(-a + \tau - \frac{\sigma_\varepsilon^2}{2}\right).
\]

The decision maker chooses capacity $M(\hat\tau^o)$ to maximize the \textit{perceived} long-run average total profit from the service system, conditional on full utilization:
\begin{equation}
\label{eq:nG_M}
\sum_{i=1}^n  G_i(M_i(\hat\tau^o)|\hat\tau^o) = n\cdot \left[p \, M(\hat{\tau}^o)
\, \lambda(\hat{\tau}^o) - s M(\hat{\tau}^o)^2\right],
\end{equation}
where $p > 0$ represents revenue per unit throughput, and $s>0$ denotes the quadratic capacity cost coefficient. The convex cost assumption ensures a well-defined interior optimum and captures increasing marginal costs of capacity expansion, consistent with organizational diseconomies of scale \citep{williamson1975markets} and labor market frictions \citep{manning2013monopsony}.
The first-order condition of \eqref{eq:nG_M} yields optimal service capacity:
$M^*(\hat{\tau}^o) = (p \, \lambda(\hat{\tau}^o))/(2s).
$
Realized profit depends on the provisioned capacity $M^*(\hat{\tau}^o)$ under the true service rate $\lambda(\tau)$, that is,
\begin{equation}
\label{eq:service_payoff}
\begin{aligned}
\sum_{i=1}^n  G_i(M(\hat\tau^o)\mid\tau)
&= n\cdot \left[p \, M^*(\hat{\tau}^o) \lambda(\tau) - s \bigl(M^*(\hat{\tau}^o)\bigr)^2 \right]\\
&= \frac{n\cdot p^2}{4s} \left[ 2\lambda(\hat{\tau}^o)\lambda(\tau) - \lambda(\hat{\tau}^o)^2 \right].
\end{aligned}
\end{equation}

Similarly, the expected payoff when untreated can be obtained by substituting $\hat\tau^o=\tau=0$ into \eqref{eq:service_payoff}.
Finally, we can obtain the SNR function as follows:
\[
\Pi(\hat{\tau}^o \mid \tau) = n\cdot C_0 \left( 2e^{\hat{\tau}^o + \tau} - e^{2\hat{\tau}^o}-1 \right).
\]
wherein
$C_0 \equiv p^2\exp\left(-2a - \sigma_\varepsilon^2\right)/(4s)$ is a constant. One can also verify that this SNR function satisfies all assumptions stated in Assumption~\ref{ass_pitauhat} and the conditions of Theorem~\ref{thm:global_convergence}.

\begin{figure}[htbp]
    \centering
    \begin{subfigure}[b]{\textwidth}
        \centering
        \includegraphics[width=0.95\textwidth]{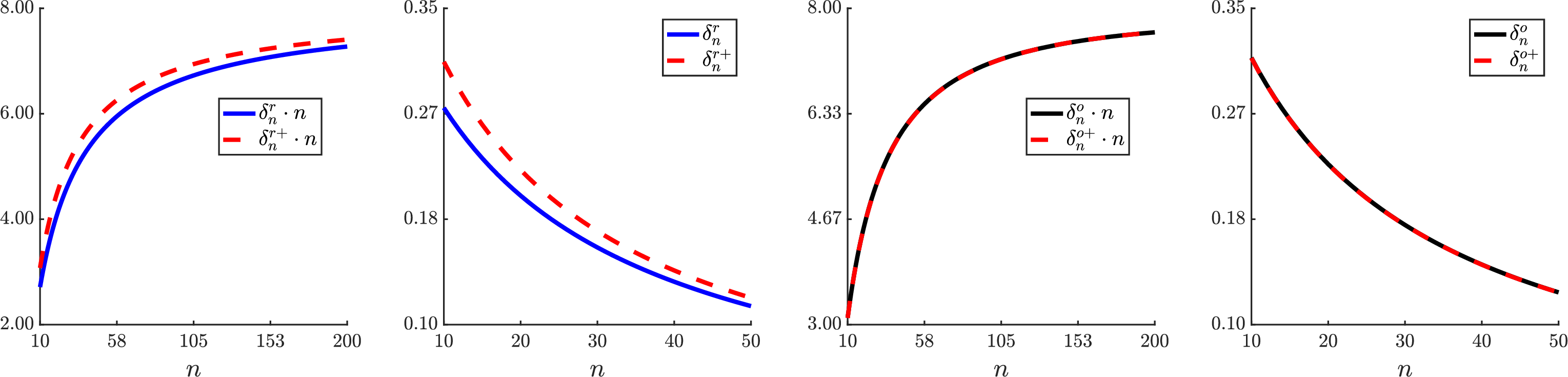}
        \caption{The convergence rate of adjustments}
        \label{fig:service_convergence}
    \end{subfigure}

    \vspace{1.5em}  

    \begin{subfigure}[b]{0.45\textwidth}  
        \centering
        \includegraphics[width=\textwidth]{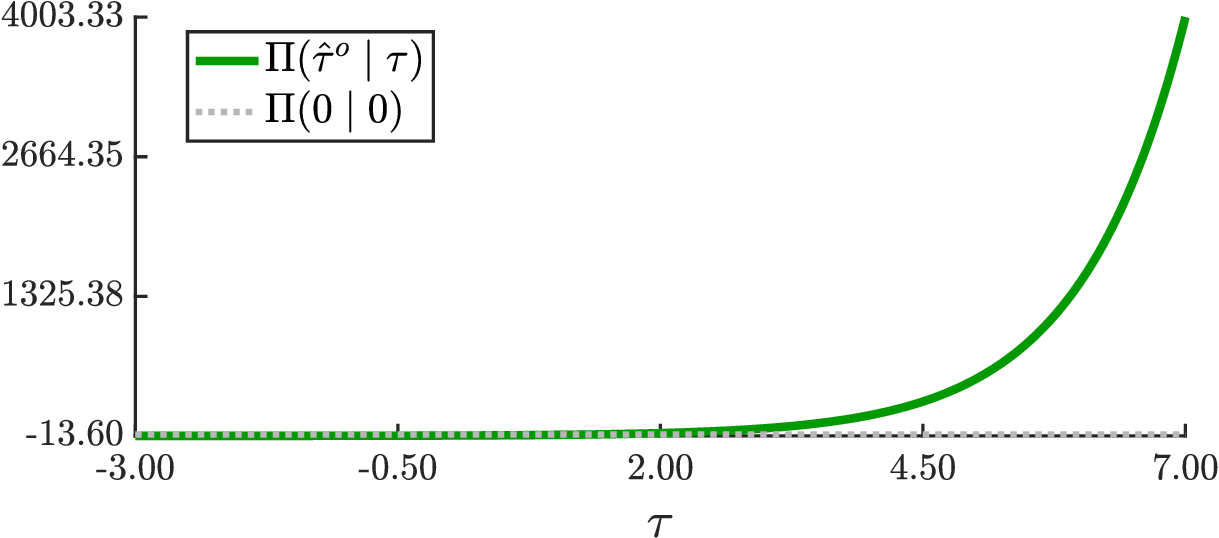}
        \caption{$\Pi(\hat{\tau}^o\mid\tau)$ given $\hat{\tau}^o=2,n=50$}
        \label{fig:Pi_wrt_tau_service}
    \end{subfigure}%
    \hspace{0.05\textwidth}
    \begin{subfigure}[b]{0.45\textwidth}  
        \centering        \includegraphics[width=\textwidth]{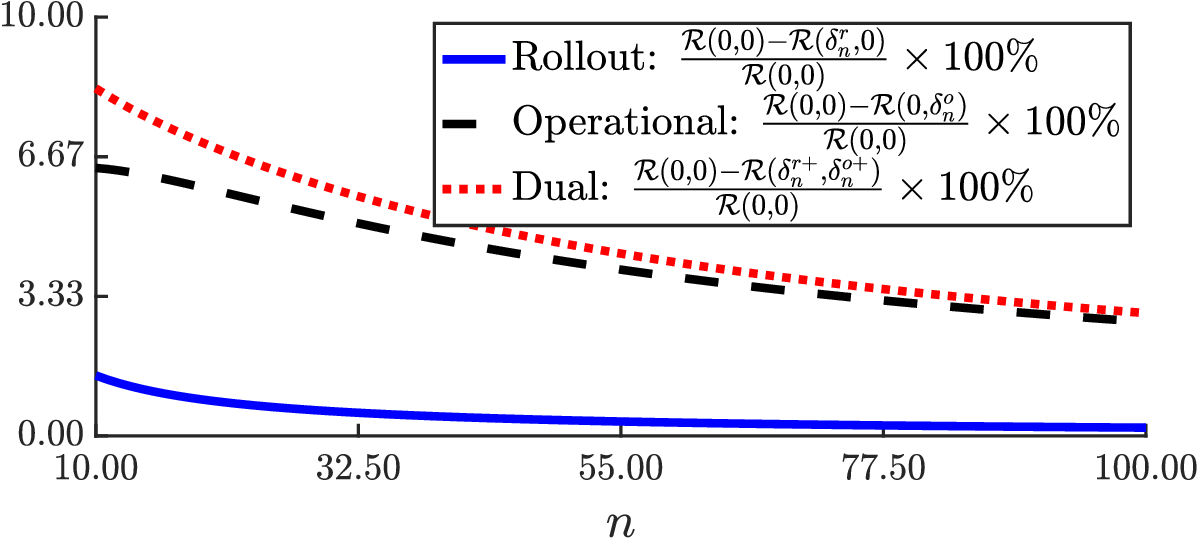}
        \caption{Regret improvement rate}
        \label{fig:service_improvement}
    \end{subfigure}

    \caption{Service Operations with Capacity Planning}
    \label{fig:service_combined}

    \vspace{1ex}
    \parbox{\textwidth}{
        \footnotesize
        \textit{Note:} The parameters are set as follows: $p=2,s=0.5,a=1,m_0=0,v_0=1,\sigma_\varepsilon=2$ and $\gamma=1$.
    }
\end{figure}

\vspace{-1em}
\paragraph{Adjustment for the rollout decision.}
The second derivative of the SNR function $\Pi(\hat{\tau}^o \mid \tau)$ with respect to $\tau$ is:
$
\Pi^{(0,2)}(\hat{\tau}^o\mid\tau) = 2nC_0 e^{\hat{\tau}^o + \tau} > 0.
$
Hence, $\Pi(\cdot \mid \tau)$ is convex in $\tau$ (see Figure~\ref{fig:Pi_wrt_tau_service}). By Proposition~\ref{prop:rollout_decision_adjustment}, the optimal adjustment for rollout decision satisfies $\delta_n^r > 0$.

The convexity arises because the value of $\tau$ (as well as its estimation error) is a logged value. When $\tau$ exceeds expectations, the actual service time is exponentially reduced. Conversely, when $\tau$ falls short, losses happen at a smaller scale. This asymmetry---amplified upside relative to downside---rewards early adoption, yielding a positive adjustment $\delta_n^r > 0$.

\vspace{-1em}
\paragraph{Adjustment for the operational decision.}
Evaluating the relevant derivatives at $\hat{\tau}^o = \tau$, we have
$
 \Pi^{(1,0)}(\tau\mid\tau) = 0,
 \Pi^{(2,0)}(\tau\mid\tau) = -2nC_0 e^{2\tau}, $ and  $
 \Pi^{(1,2)}(\tau\mid\tau)=-2nC_0 e^{2\tau}.
$ By Corollary~\ref{corr:variable_adjustment_direction}, the optimal adjustment for the operational decision satisfies $\delta_n^o < 0$, as verified in Figure~\ref{fig:service_convergence}.

\vspace{-1em}
\paragraph{Dual adjustments.}
Numerical analysis confirms that this example satisfies condition (b) in Case I of Proposition~\ref{prop:complementary}, yielding $\delta_n^{r+} > \delta_n^r > 0$ (see Figure~\ref{fig:service_convergence}).

\subsection{Promotion and Pricing with Generalized Linear Demand}\label{sec:pricing_example}
In this example, we consider experiments that evaluate whether a treatment can improve the expected demand, followed by rollout and downstream pricing decisions. In this example, we consider a generalized linear demand specification with additive uncertainty:
\begin{equation*}
    Q(a,b,p,\lambda)=\left[1+\lambda(a-bp+\varepsilon)\right]^{1/\lambda},
\end{equation*}
where $a$ is the baseline market potential, $b$ is the price sensitivity, $p$ is the price, $\varepsilon$ is a random shock of zero mean independent of treatment, and $\lambda$ is a hyperparameter. When $\lambda\rightarrow1$, it becomes a linear demand; when $\lambda\rightarrow0$, it becomes a log-linear demand. Without loss of generality, we focus on these two degenerated cases. If the treatment is implemented, the market potential becomes $a+\tau$ and the demand becomes $Q(a+\tau,b,p,\lambda)$.

\begin{figure}[htbp]
    \centering

    \begin{subfigure}[b]{0.9\textwidth}
        \centering
        \includegraphics[width=\textwidth]{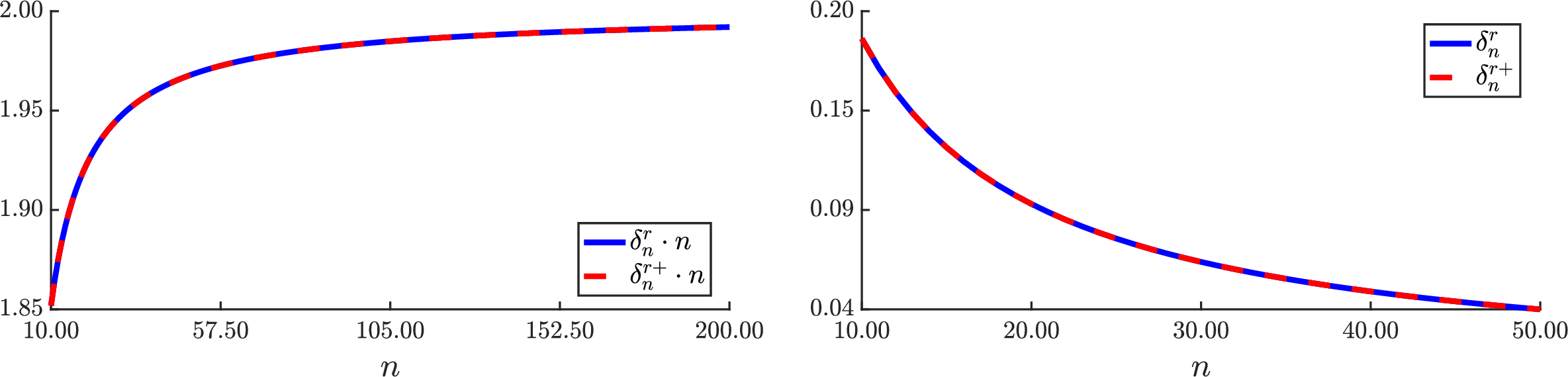}
        \caption{The convergence rate of rollout adjustments}
        \label{fig:convergence_log_pricing}
    \end{subfigure}%
    \vspace{1.5em}

   \begin{subfigure}[b]{0.43\textwidth}  
        \centering
        \includegraphics[width=\textwidth]{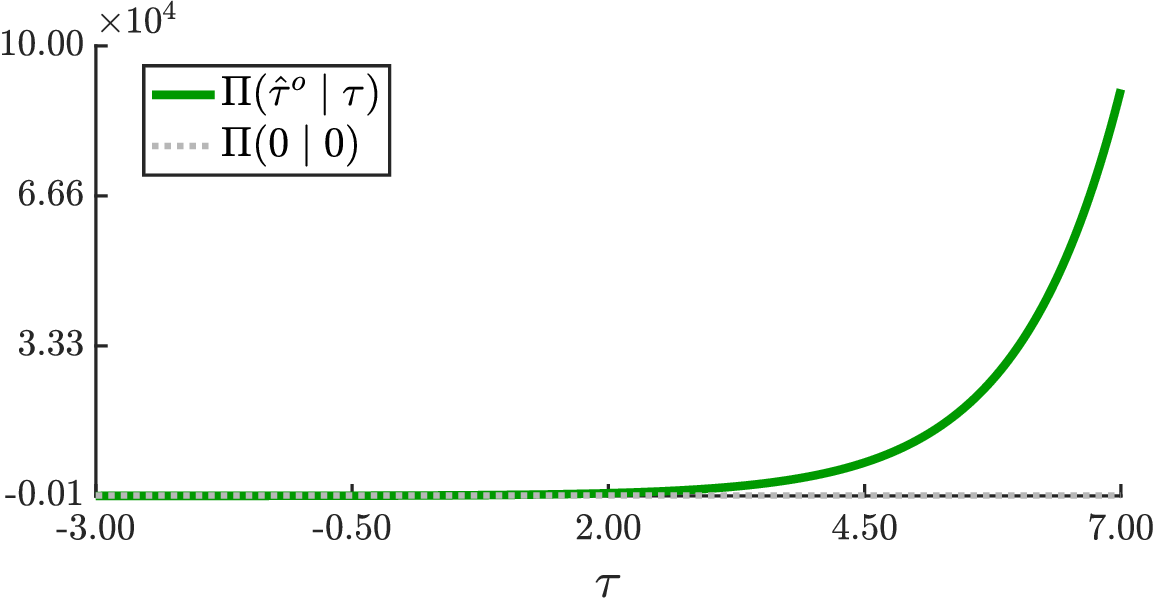}
        \caption{$\Pi(\hat{\tau}^o\mid\tau)$ given $\hat{\tau}^o=2, n=50$}
        \label{fig:Pi_logdemand}
    \end{subfigure}%
    \hspace{0.03\textwidth}
    \begin{subfigure}[b]{0.43\textwidth}
        \centering
        \includegraphics[width=\textwidth]{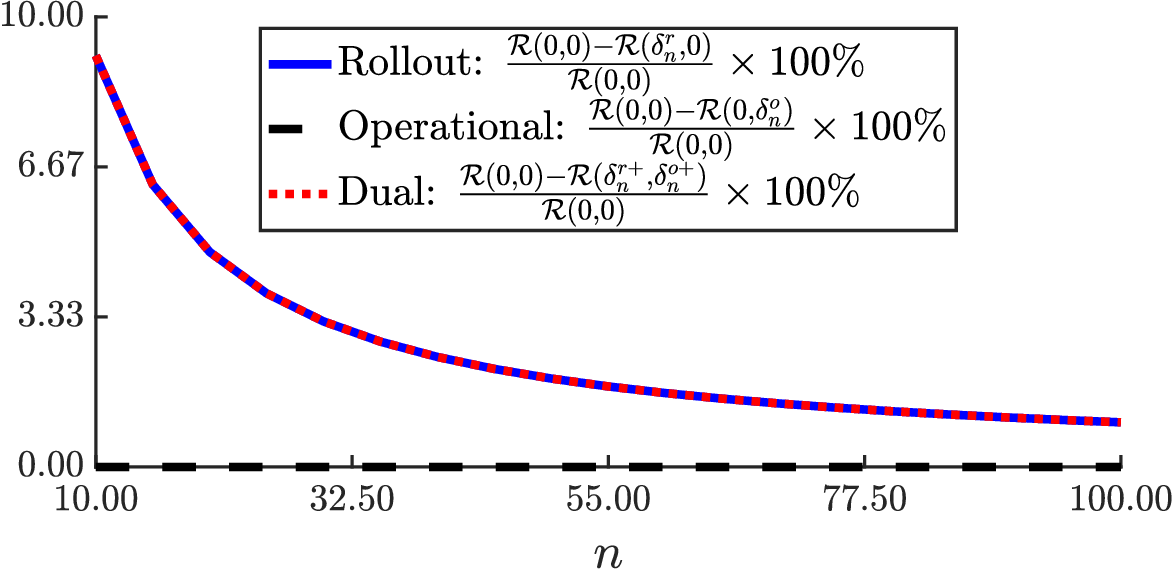}
        \caption{Regret improvement rate}
        \label{fig:exp_improvement}
    \end{subfigure}  \caption{Pricing with Log-Linear Demand}
\label{fig:pricing}
\vspace{1ex} 
    \parbox{\textwidth}{
        \footnotesize 
        \textit{Note:} The parameters are set as follows: $m_0=0,v_0=5,\sigma_\varepsilon=1,a=b=1$, and $\gamma=1$.}
\end{figure}

The firm estimates the treatment effect as $\hat{\tau}^o$ and solves for the price $p^*(\hat{\tau}^o)$ that maximizes the perceived expected total revenue, condition on implementation:
$$
\sum_{i=1}^n G_i(p|\hat\tau^o)=n \cdot p\cdot\mathbb{E}_{\varepsilon}[Q(a+\hat\tau^o,b,p,\lambda)].
$$
The first-order condition yields:
$$
p_\lambda^*(\hat\tau^o)=
\begin{cases}
    (1+a+\hat\tau^o)/(2b)&\text{if}\quad \lambda\rightarrow 1;\\
    1/b&\text{if}\quad \lambda\rightarrow 0.
\end{cases}
$$
Accordingly, the SNR function is given by
$$
\Pi_\lambda(\hat\tau^o\mid\tau)=
\begin{cases}
    n\cdot \frac{2(a+1)\tau-(\hat{\tau}^o)^2+2\tau \hat{\tau}^o}{4b}&\text{if}\quad \lambda\rightarrow 1;\\
    \frac{n}{b}\left[\exp\!\left(a - 1 + \tau + \frac{\sigma_\varepsilon^2}{2}\right)-\exp\!\left(a - 1 +\frac{\sigma_\varepsilon^2}{2}\right)\right]&\text{if}\quad \lambda\rightarrow 0.
\end{cases}
$$
\paragraph{Linear demand ($\lambda\rightarrow1$).} The SNR function is linear in the treatment effect, so there is no adjustment for the rollout decision (i.e., $\delta_n^r = 0$). In addition, the SNR is quadratic and concave in $\hat{\tau}^o$ with constant curvature, so there is also no adjustment for the pricing decision (i.e., $\delta_n^o = 0$).
\vspace{-1em}
\paragraph{Log-linear demand ($\lambda\rightarrow0$).} The SNR function is strictly convex in $\tau$, so the firm should be aggressive in rolling out the intervention and thus the optimal adjustment for the rollout decision is positive (i.e., $\delta_n^r>0$). As the SNR function is independent of $\hat{\tau}^o$, all higher-order derivatives with respect to the estimator vanish. Therefore, there is no adjustment for the pricing decision (i.e., $\delta^o_n = 0$) and thus for the rollout decision $\delta_n^{r+} = \delta_n^r$. The results are illustrated in Figure \ref{fig:pricing}.

\section{Conclusion}
\label{sec:conclusion}

This paper studies how firms should convert noisy experimental evidence into high-quality post-experiment decisions when implementation is followed by downstream operational re-optimization. We propose a simple Predict-Adjust-Then-Rollout-Optimize (PATRO) approach: keep standard causal estimation, but intentionally adjust the estimated treatment effect before it is used for (i) the rollout decision and (ii) the operational optimization. In a Bayesian framework with a Gaussian posterior, these adjustments correspond to choosing posterior quantiles (equivalently, additive shifts from the posterior mean) to minimize prior expected regret.

Our analysis delivers the following takeaways. First, post-experiment decision quality generally requires two distinct corrections: the rollout threshold and the downstream optimizer respond to different (and often asymmetric) regret components, so a single “significant-and-plug-in” estimate is not economically calibrated in small samples. Second, the direction and interaction of these corrections are governed by the geometry of the downstream value function: curvature in the true effect determines whether rollout should be conservative or aggressive, while cross-curvature (“2D skewness”) determines whether downstream decisions should be biased up or down—and, in two-stage settings, the two adjustments can be substitutes or complements, so they must be chosen jointly rather than stage-by-stage. Third, PATRO shows that one can capture (nearly) Bayes-optimal performance with a transparent, operationally lightweight rule: fixed posterior-quantile plug-ins with ex ante–computed additive shifts, supported by first-order characterizations, $O(n^{-1})$ scaling, and a convergent alternating algorithm—yielding a practical bridge between standard PTO pipelines and fully Bayes decision rules.

Our baseline analysis assumes a homogeneous treatment effect~\(\tau\) across units. The framework can be generalized to heterogeneous effects by letting each unit-specific effect \(\tau_i\) be drawn i.i.d.\ from \(N(\tau,\sigma_\tau^2)\), while \(\tau\) retains its original prior. This formulation is equivalent to maintaining the constant-effect potential outcome model in~\eqref{eq:potential}, except that the treated and control groups now exhibit different noise variances: the control variance remains known from historical data, whereas the treated variance is larger and unknown. Assigning the latter an inverse-Gamma prior introduces additional analytical complexity, but we conjecture that the qualitative insights—especially about the structure of optimal rollout and operational policies—remain unchanged.


\singlespacing
\bibliographystyle{plainnat}
\bibliography{ref}

\end{document}